%% file: main_ieee.tex
\documentclass[conference]{IEEEtran}
\IEEEoverridecommandlockouts

\usepackage{cite}
\usepackage{amsmath,amssymb,amsfonts}
\usepackage{algorithmic}
\usepackage{graphicx}
\usepackage{textcomp}
\usepackage{xcolor}
\usepackage[hyphens]{url}
\usepackage{booktabs}

\def\BibTeX{{\rm B\kern-.05em{\sc i\kern-.025em b}\kern-.08em
    T\kern-.1667em\lower.7ex\hbox{E}\kern-.125emX}}

\newcommand{\secref}[1]{Sec.~\ref{#1}}
\newcommand{\figref}[1]{Fig.~\ref{#1}}
\newcommand{\tabref}[1]{Table~\ref{#1}}
\newcommand{\cmark}{\ding{51}} 
\newcommand{\xmark}{\ding{55}} 
\usepackage{enumitem}
\usepackage{indentfirst}
\usepackage{fancyhdr} 
\usepackage{pifont}
  
\usepackage{array}
\newcolumntype{L}[1]{>{\raggedright\let\newline\\\arraybackslash\hspace{0pt}}m{#1}}
\newcolumntype{C}[1]{>{\centering\let\newline\\\arraybackslash\hspace{0pt}}m{#1}}
\newcolumntype{R}[1]{>{\raggedleft\let\newline\\\arraybackslash\hspace{0pt}}m{#1}}

\usepackage[bookmarks=true,breaklinks=true,letterpaper=true,colorlinks,citecolor=blue,linkcolor=blue,urlcolor=blue]{hyperref}

\begin{document}

\pdfpagewidth=8.5in
\pdfpageheight=11in

\newcommand{\iscasubmissionnumber}{1727}

\pagenumbering{arabic}

\title{Efficient 3D Gaussian Splatting with Axis-Shared Rasterization and Order-independent Transmittance}

\author{
    \IEEEauthorblockN{
        Zhican Wang\IEEEauthorrefmark{1},
        Guanghui He\IEEEauthorrefmark{1}\thanks{This work was supported by the National Natural Science Foundation of China under Grant U25B2057 and Grant 92464302. Corresponding author: Guanghui He.},
        Lingjun Gao\IEEEauthorrefmark{3},
        Dantong Liu\IEEEauthorrefmark{2},
        Shell Xu Hu\IEEEauthorrefmark{4}, \\
        Chen Zhang\IEEEauthorrefmark{1},
        Zhuoran Song\IEEEauthorrefmark{1},
        Nicholas Lane\IEEEauthorrefmark{2}, and
        Hongxiang Fan\IEEEauthorrefmark{3}
    }
    \vspace{0.15cm}
    \IEEEauthorblockA{\IEEEauthorrefmark{1}Shanghai Jiao Tong University  \{wang\_zhican, guanghui.he, chenzhang, songzhuoran\}@sjtu.edu.cn}
    \IEEEauthorblockA{\IEEEauthorrefmark{2}University of Cambridge  liudt921115@gmail.com, ndl32@cam.ac.uk}
    \IEEEauthorblockA{\IEEEauthorrefmark{3}Imperial College London  \{lingjun.gao24, hongxiang.fan\}@imperial.ac.uk}
    \IEEEauthorblockA{\IEEEauthorrefmark{4}Samsung AI  shell.hu@samsung.com}
}





\maketitle
\thispagestyle{plain}
\pagestyle{plain}

\begin{abstract}
3D Gaussian Splatting (3DGS) has emerged as a powerful technique for novel view synthesis, combining high-quality reconstruction with efficient rendering. It has been widely adopted in domains such as AR/VR, robotics, and autonomous driving. However, achieving real-time performance on resource-constrained platforms remains challenging due to strict power and area budgets. Prior accelerators improve hardware performance but still overlook key inefficiencies, including insufficient rasterization efficiency, poor sorting scalability, and pipeline imbalance. This paper presents an architecture--algorithm co-design to address these challenges. First, we propose \textit{axis-shared rasterization}, which precomputes and reuses common terms along the X- and Y-axes, reducing multiply-and-accumulate (MAC) operations by up to $38\%$ while preserving high parallelism. Second, we develop a novel \textit{order-independent transmittance} method that removes the need for explicit sorting by leveraging a lightweight Multilayer Perceptron (MLP) to directly approximate the transmittance of each Gaussian, enabling efficient $\alpha$ blending with negligible quality loss. Third, we design a \textit{unified reconfigurable PE array} that supports both rasterization and MLP inference, sustaining high utilization without costly sorting hardware. Our experiments demonstrate that our design preserves rendering quality while achieving a $1.33\sim 1,88\times$ speedup over the state-of-the-art 3DGS accelerators. Our code is open source at \url{https://github.com/WangZhican/ISCA26\_3DGS\_Acc}.

\end{abstract}

\input{./text/0_intro}

\input{./text/1_background}

\input{./text/2_rasterization}
\input{./text/3_algorithm}

\input{./text/4_hardware}

\input{./text/5_evaluation}
\input{./text/7_related_work}

\input{./text/6_conclusion}

\newpage
\bibliographystyle{IEEEtranS}
\bibliography{sample-base}

\end{document}

%% file: text/0_intro.tex
\section{Introduction}
\label{sec:intro}

3D Gaussian Splatting (3DGS)~\cite{kerbl20233d} has emerged as a prominent technique for novel view synthesis, offering both high-quality reconstruction and efficient rendering performance. It has been widely adopted across diverse domains, including robotics~\cite{zhu20243d}, augmented and virtual reality (AR/VR)~\cite{tu2024fast, zhai2025splatloc}, and autonomous driving~\cite{khan2024autosplat}. In contrast to neural radiance fields (NeRF)~\cite{mildenhall2021nerf}, which implicitly represent 3D scenes using neural networks, 3DGS explicitly encodes scenes as a large set of 3D Gaussians with learnable positions, sizes, shapes, colors, and opacities. Owing to its lower algorithmic complexity, 3DGS achieves significantly faster rendering performance than NeRF, making it particularly suitable for interactive applications. However, achieving real-time 3DGS rendering on resource-constrained platforms, such as edge GPUs, remains a significant challenge. For instance, on an NVIDIA Jetson Orin Nano edge GPU~\cite{nvidia_orin_nano}, we observe only approximately $20$ frames per second on the MipNeRF-360 dataset~\cite{barron2022mip}. The stringent power and area constraints of AR/VR edge devices further exacerbate the difficulty of deploying 3DGS in practice.

\begin{figure*}[ht]
\centering
\includegraphics[width=176mm]{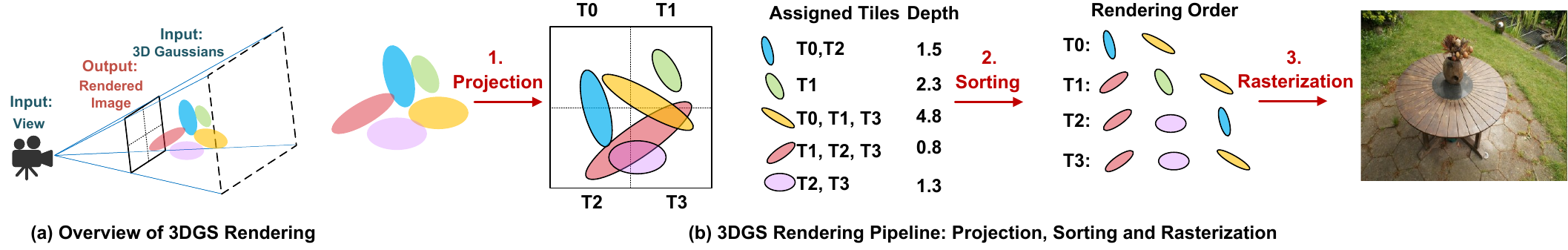}
\vspace{-10pt}
\caption{The rendering process of 3D Gaussian Splatting.}
\label{fig:3DGS}
\vspace{-10pt}
\end{figure*}


\begin{table}[t]
\centering
\setlength{\tabcolsep}{4pt}
\small
\caption{Comparison with related 3DGS accelerators.}
\label{tab:sota_comp}\scalebox{0.8}{
\begin{tabular}{lcccc}
\toprule
 & GSCore~\cite{lee2024gscore} 
 & Meta.~\cite{10.1145/3669940.3707227} 
 & GBU~\cite{ye2025gbu} 
 & Ours \\
\midrule
Ras. MAC Reduction 
  & \xmark & \xmark & \cmark & \cmark \\

Parallel Rasterization 
  & \cmark & \cmark & \xmark & \cmark \\

Sorting Implementation
  & Hier. Bitonic
  & Hier. Bitonic 
  & GPU-based 
  & Replaced \\
\bottomrule
\end{tabular}}
\vspace{-6pt}
\end{table}

Given a camera pose as input, 3DGS renders a scene represented by 3D Gaussians (ellipsoids) into a final image, as illustrated in \figref{fig:3DGS}(a). The rendering pipeline comprises three essential steps: projection, sorting, and rasterization (details are provided in \secref{sec:background}). Profiling results, presented in \figref{fig:profile} (left), indicate that rasterization and sorting dominate the overall latency, consistent with prior studies~\cite{10.1145/3669940.3707227, lee2024gscore, ye2025gbu, 10.1145/3695053.3731003}. Existing approaches, including GSCore and MetaSapiens~\cite{lee2024gscore, 10.1145/3669940.3707227}, enhance efficiency via Gaussian count reduction but neglect the inherent computational redundancy in rasterization. GBU~\cite{ye2025gbu} decreases rasterization MACs using spatial transformations and sequential differential computation; however, this introduces inter-pixel dependencies, thereby compromising pixel-level parallelism. Another work, Lumina~\cite{10.1145/3695053.3731003}, reduces the computation of sorting and rasterization by exploiting the similarity between consecutive frames; however, it is applicable only to moving-view scenarios, whereas our method does not rely on this assumption. Moreover, in these methods, the overhead of sorting and the possibility of replacing it with a more efficient alternative remain insufficiently explored. A comparison with related 3DGS accelerators is presented in \tabref{tab:sota_comp}. These limitations motivate a more in-depth analysis of rasterization and sorting, culminating in the two key challenges outlined below:

\begin{itemize}[leftmargin=*]
\item \textit{\textbf{Challenge-1: Computational redundancy in rasterization.}}  
Most existing implementations adopt a pixel-wise mapping strategy, where each pixel is rasterized independently by a separate processing element (PE) in an accelerator or a CUDA core in a GPU. While this approach offers high parallelism, it ignores the substantial overlap in intermediate computations shared across neighboring pixels. As a result, many common terms are redundantly recalculated along rows and columns, leading to unnecessary multiply-and-accumulate (MAC) operations. This redundancy significantly increases latency or PE overhead, making rasterization inefficient. A deeper analysis is provided in \secref{sec:challenge_red}.

\item \textit{\textbf{Challenge-2: Sorting scalability and pipeline imbalance.}}  
Hardware implementations of sorting often rely on parallel sorting modules~\cite{ionescu1997optimizing}, whose area and cost scale rapidly with input parallelism. Moreover, sorting and rasterization have inherently different computational complexities, $O(N\log^2 N)$ for sorting versus $O(N)$ for rasterization, posing challenges for maintaining pipeline balance. This imbalance is further exacerbated by large variations in the number of Gaussians per tile, which span up to two orders of magnitude in our profiling. Consequently, fixed-parallelism sorting modules either lead to resource underutilization on small workloads or cause rasterization bottlenecks for large workloads. A detailed analysis is provided in \secref{sec:challenge_sort}.
\end{itemize}

To address \textit{\textbf{Challenge-1}}, we propose an \textit{axis-shared rasterization} technique that eliminates redundant computations within each tile. The core idea is to compute common intermediate terms along the X- and Y-axes only once and then share these terms across all processing elements (PEs) within the tile. Each PE can then complete rasterization by simply combining the shared terms. This design reduces the number of MAC operations by $38\%$ while preserving high parallelism. As a result, it enables either lower latency under the same PE resources or reduced PE overhead with equivalent latency. Building on this principle, we further design a highly efficient PE array architecture for rasterization.

To address \textit{\textbf{Challenge-2}}, we adopt a hardware-algorithm co-design approach. On the algorithmic side, we re-examine the role of sorting in Gaussian Splatting, which traditionally enforces depth ordering for transmittance computation and ultimately determines the decay factor for color blending. Our key insight is that the decay factor can be computed directly without explicit sorting. By recognizing the analogy between 3DGS and image composition~\cite{porter1984compositing}, and inspired by order-independent transparency techniques~\cite{carpenter1984buffer, mcguire2013weighted}, we develop a novel \textit{order-independent transmittance} (OIT) method tailored to 3DGS. This method, co-designed with the hardware, leverages a lightweight MLP to predict decay factors, mitigating workload imbalance while preserving image quality. On the hardware side, we introduce a unified and reconfigurable PE array that supports both rasterization and MLP inference, thereby removing the need for a costly sorting engine and sustaining consistently high utilization.

In conclusion, we make the following contributions:
\begin{itemize}[leftmargin=*]
    \item We identify two previously overlooked challenges in accelerating 3D Gaussian Splatting, \textit{(i)} computational redundancy in rasterization and \textit{(ii)} scalability and pipeline imbalance issues caused by sorting (\secref{sec:challenge}).
    
    \item We propose an \textit{axis-shared rasterization} technique and a dedicated hardware design, eliminating redundant computations and reducing MAC operations by $38\%$ (\secref{sec:axes}).
    
    \item We develop a novel \textit{order-independent transmittance} method that bypasses explicit sorting and enables efficient decay factor prediction with negligible quality loss (\secref{sec:alg}).
    
    \item We design a \textit{unified and reconfigurable hardware accelerator} that supports both rasterization and MLP inference, achieving real-time 3DGS rendering with consistently high utilization (\secref{sec:hardware}).
\end{itemize}

%% file: text/1_background.tex
\section{Background and Motivation}

\subsection{3D Gaussian Splatting}
\label{sec:background}
\textbf{3DGS parameters.} 3DGS represents a 3D scene as a collection of Gaussians. For a given camera view, an image is rendered by splatting these Gaussians into 2D space following the standard 3DGS pipeline. Each Gaussian is defined by Equation~(\ref{eq:gs_model}), where $p'$ denotes the 3D coordinates. In total, 59 parameters are required to describe each Gaussian: \textit{i)} the mean (position) $\mu'$ (3 parameters); \textit{ii)} the size and shape, represented by the covariance matrix $\Sigma'$, which is determined by the scale $s$ (3 parameters) and rotation $q$ (4 parameters); \textit{iii)} the opacity factor $o$ (1 parameter); and \textit{iv)} the view-dependent color, parameterized by spherical harmonics (SH) coefficients (16$\times$3 = 48 parameters).

\textbf{Rendering steps.} The overall rendering process takes the camera parameters, pose, and 3D Gaussians as input, and synthesizes a final image as output, as shown in \figref{fig:3DGS}(a). The rendering pipeline consists of three main steps—projection, sorting, and rasterization—as illustrated in \figref{fig:3DGS}(b). \textit{For projection}, based on the camera parameters and pose, 3D Gaussians are projected into 2D Gaussians. Specifically, the 3D mean ($\mu'$) and covariance ($\Sigma'$) are projected into a 2D mean $\mu$ ($2\times1$) and a 2D covariance $\Sigma$ ($2\times2$). The depth ($d$) of each 3D Gaussian relative to the camera is also calculated (e.g., $0.8$–$4.8$ in the figure). In addition to spatial attributes, color is represented by an RGB vector ($3\times1$), computed from the SH coefficients and camera pose. Rendering is performed at a fixed tile granularity of $16\times16$ pixels. After projection, each 2D Gaussian is mapped to the tiles it overlaps. For example, in the toy case shown, the blue Gaussian overlaps tiles $T0$ and $T1$, while the green Gaussian overlaps only $T1$. \textit{For sorting}, since the relative depth order determines occlusion and affects $\alpha$-blending, each tile sorts its intersecting Gaussians in ascending depth order (near to far). As shown in \figref{fig:3DGS}, the correct order for tile $T0$ is the blue Gaussian followed by the yellow. \textit{For rasterization}, the pipeline computes each Gaussian’s contribution to every pixel within a tile, ultimately synthesizing the final image. This process consists of $\alpha$-computation followed by $\alpha$-blending. The $\alpha$ value is computed using Equation~(\ref{eq:a_comp}), where $\mathbf{p}$ ($2\times1$) denotes the pixel position, $\mu$ and $\Sigma$ are the projected 2D Gaussian mean and covariance, and $o$ is the opacity factor. Based on the computed $\alpha$ values, $\alpha$-blending determines the final pixel color $C$ as defined in Equation~(\ref{eq:a_blending}) (left), where $T_i$ denotes the accumulated transmittance, $i$ indexes the sorted Gaussians, and $c_i$ denotes each Gaussian’s RGB color. The transmittance $T_i$ is computed from all $i-1$ preceding Gaussians, as given in Equation~(\ref{eq:a_blending}) (right).

\begin{equation}
\label{eq:gs_model}
G(\mathbf{p'}) = e^{-\frac{1}{2} (\mathbf{p'} - \boldsymbol{\mu'})^\mathrm{T} \boldsymbol{\Sigma'}^{-1} (\mathbf{p'} - \boldsymbol{\mu'})},
\end{equation}
\begin{equation}
\label{eq:a_comp}
\alpha = o \cdot e^{ -\frac{1}{2} (\mathbf{p} - \boldsymbol{\mu})^\mathrm{T} \boldsymbol{\Sigma}^{-1} (\mathbf{p} - \boldsymbol{\mu}) },
\end{equation}
\begin{equation}
\label{eq:a_blending}
C = \sum_{i=1}^{N} T_i \alpha_i c_i, \quad T_i = \prod_{j=1}^{i-1} (1 - \alpha_j)
\end{equation}

 \begin{figure}[ht]
 \vspace{-15pt}
\centering
\includegraphics[width=84mm]{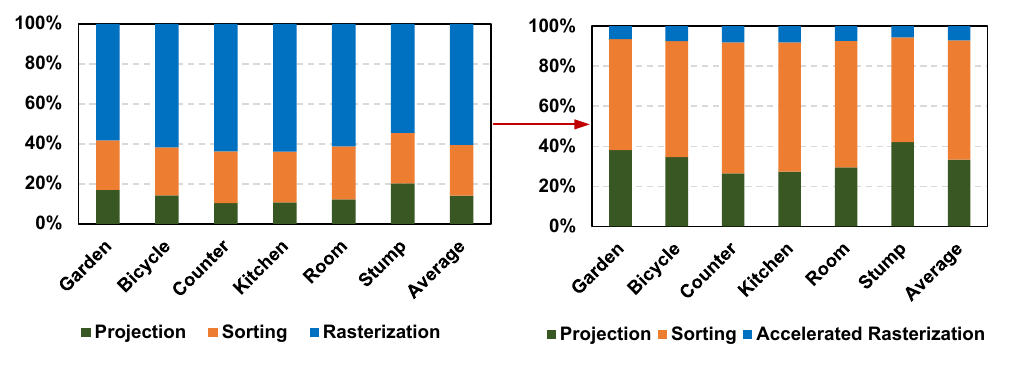}
\vspace{-10pt}
\caption{3DGS latency breakdown on GPU (left) and GPU with hardware accelerated rasterization(right).}
\label{fig:profile}
\vspace{-5pt}
\end{figure} 

\textbf{Profiling and analysis.} We conducted a detailed profiling of the three rendering steps on the NVIDIA Jetson Orin Nano GPU using the MipNeRF-360 dataset~\cite{barron2022mip}, as shown in \figref{fig:profile} (left). The results indicate that projection, sorting, and rasterization account for $14.2\%$, $25.3\%$, and $60.5\%$ of the total latency, respectively. Since sorting and rasterization together constitute nearly $90\%$ of the latency, our work primarily targets the acceleration of these two steps. An analysis of the MAC count per Gaussian within the rasterization step shows that $\alpha$-computation requires $8$ multiplications, $4$ additions, and $1$ exponential operation, whereas $\alpha$-blending requires $5$ multiplications and $4$ additions. This reveals that $\alpha$-computation is the most MAC-intensive operation, motivating our focus on optimizing it.


\subsection{Challenge Analysis and Motivation}
\label{sec:challenge}

\subsubsection{ \textit{\textbf{Challenge-1: Computational redundancy in rasterization.}}}
\label{sec:challenge_red}
\begin{figure}[hbt]
\centering
\includegraphics[width=84mm]{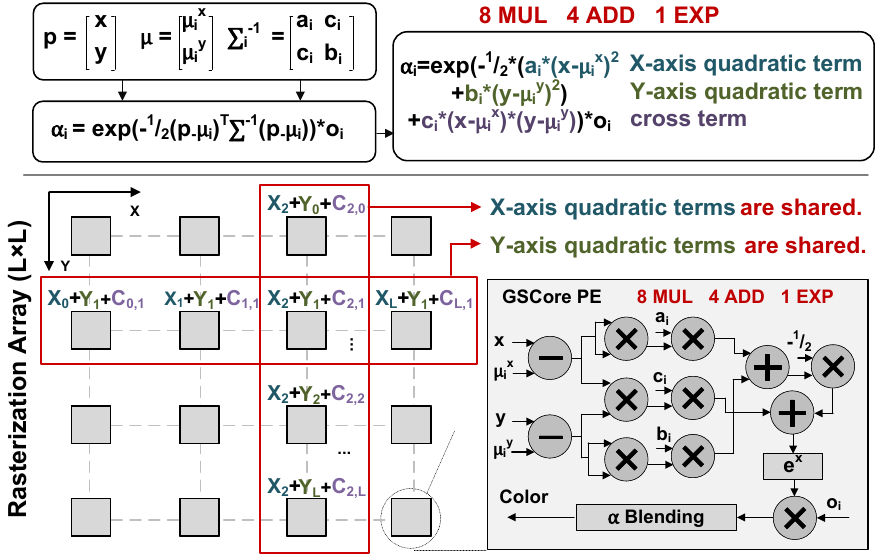}
\vspace{-5pt}
\caption{The sources of redundant computing in rasterization.}
\label{fig:redundancy}
\vspace{-10pt}
\end{figure}

As demonstrated in profiling, rasterization is the most time-consuming component in 3DGS.
Within the rasterization pipeline, $\alpha$-computation is the most MAC-intensive process, involving cascaded matrix-vector multiplications.
The formula expansion for $\alpha$ computing is shown in \figref{fig:redundancy} (top), where $\mu_{i}^{x}$ and $\mu_{i}^{y}$ denote the center of the $i$-th Gaussian, $x$ and $y$ denote the coordinates of a pixel.
The conic matrix ($\Sigma^{-1}$), which is defined as the inverse of 2D Gaussian's covariance matrix, is parameterized by $a_i$, $b_i$, and $c_i$. According to the formulation, it requires $8$ multiplications (MUL), $4$ additions (ADD), and $1$ exponential operation (EXP). Conventionally, prior work ~\cite{lee2024gscore} designs a PE array for rasterization, where the computation for each pixel is mapped to one PE.
In these designs, the PE structure reflects the theoretical MAC count derived from the $\alpha$ computation formulation, as shown in ~\figref{fig:redundancy} (bottom), with registers omitted for simplicity.

To illustrate the source of redundancy, we decompose the exponent in the $\alpha$ computation into three components, as shown in \figref{fig:redundancy} (top): \textit{(i)} the X-axis quadratic term, representing the squared distance between the pixel and the Gaussian center along the X-axis; \textit{(ii)} the Y-axis quadratic term, representing the squared distance along the Y-axis; and \textit{(iii)} the cross term, capturing the interaction between the X and Y coordinates. When mapping pixel computations onto the PEs of the rasterization array (\figref{fig:redundancy}, bottom), clear spatial redundancy emerges: within any row of pixels, the Y-axis quadratic terms remain identical across PEs, while within any column, the X-axis quadratic terms are shared. This analysis highlights substantial redundancy in both the X- and Y-axis quadratic terms. Yet conventional PE designs recompute these terms independently for every pixel, missing the opportunity to exploit this redundancy for computational savings, as illustrated in \figref{fig:redundancy} (bottom right).

To avoid redundant computation, our key idea is to redesign the entire dataflow by precomputing the axis-shared terms. Each PE then performs only a simple combination of these precomputed values, thereby significantly reducing complexity while maintaining parallelism. This axis-shared rasterization is implemented through a dedicated hardware architecture~(\secref{sec:axes}) to improve the compute efficiency.

\subsubsection{\textit{\textbf{Challenge-2: Sorting scalability and pipeline imbalance issues.}} }
\label{sec:challenge_sort}
\begin{figure}[ht]
\centering
\includegraphics[width=84mm]{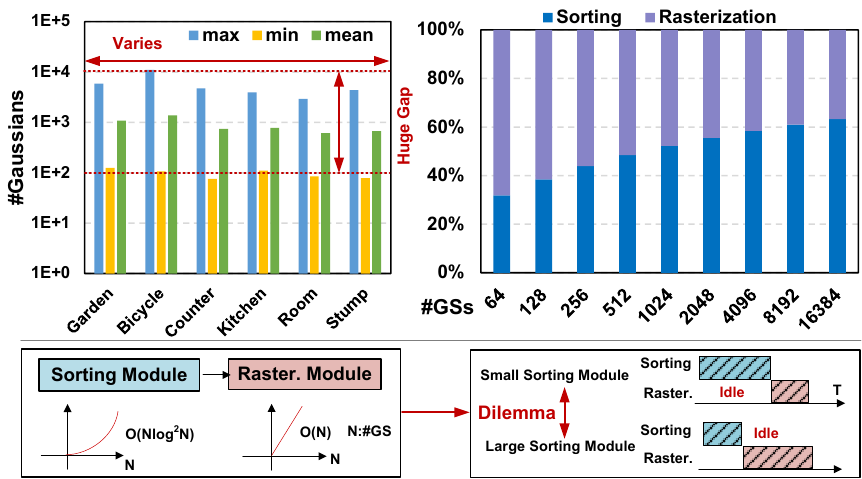}
\vspace{-5pt}
\caption{The number of Gaussians varies across tiles and scenes and leads to pipeline imbalance issue.}
\label{fig:gs_count}
\vspace{-5pt}
\end{figure} 

According to Amdahl’s Law~\cite{rodgers1985improvements}, if only rasterization is accelerated while other steps remain on the GPU, sorting inevitably becomes the primary bottleneck, as shown in \figref{fig:profile} (right). To quantitatively assess the impact of sorting on per-tile rendering, we profiled the MipNeRF-360 dataset~\cite{barron2022mip}. We employed a trained Gaussian checkpoint at 7k iterations and used the validation set to measure the variation in Gaussian counts across tiles and scenes. As shown in~\figref{fig:gs_count} (top left), the per-tile Gaussian count ranges from approximately $80$ to over $10,000$, spanning more than two orders of magnitude. This variation occurs both within tiles of a single scene and across different scenes, presenting a fundamental challenge for efficient hardware design.

We further profile the latency of rasterization and sorting across different Gaussian counts per tile, following the hardware architecture of \cite{lee2024gscore}, as shown in \figref{fig:gs_count} (top right). When the Gaussian count is small, the sorting latency can be less than half that of rasterization, whereas for large counts, it can grow to nearly twice as high. This trend can be explained through complexity analysis. Let $N$ denote the number of Gaussians mapped to a tile. Sorting typically incurs $O(N\log^2 N)$ complexity, as in bitonic sort~\cite{ionescu1997optimizing}, with hardware area scaling as $k\log^2 k$ under input parallelism $k$, leading to rapidly increasing overhead as $k$ grows. In contrast, rasterization is a MAC-dominated operation with linear complexity $O(N)$. Due to this heterogeneity, sorting and rasterization cannot be efficiently unified into a single engine. Moreover, the disparity in computational complexity, combined with the large variance in Gaussian counts across tiles, inherently causes pipeline imbalance. As illustrated in the schedule of \figref{fig:gs_count} (bottom), under a fixed area budget, a small sorting module becomes the bottleneck for tiles with many Gaussians, stalling rasterization and reducing utilization. Conversely, a larger sorting module with higher overhead shifts the bottleneck to rasterization for tiles with fewer Gaussians. Thus, designing an architecture that achieves balanced performance across diverse scenes and tile workloads remains a significant challenge.

To comprehensively address this challenge, we adopt an algorithm–hardware co-design strategy. At the algorithmic level, we revisit the role of sorting and entirely replace it with an order-independent transmittance method (\secref{sec:alg}). It exploits the high throughput of the PE array through reconfigurability, enabling uniform support for both transmittance computation and rasterization (\secref{sec:hardware}). Consequently, regardless of the number of Gaussians per tile, the unified hardware sustains consistently high utilization.

%% file: text/2_rasterization.tex
\section{Axis-Shared Rasterization}
\label{sec:axes}
\begin{figure}[ht]
 \vspace{-5pt}
\centering
\includegraphics[width=84mm]{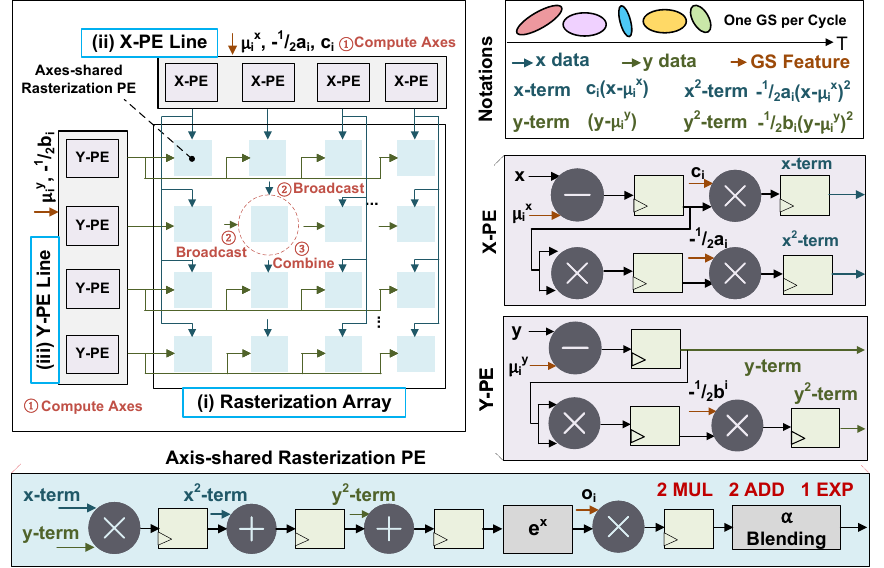}
\vspace{-5pt}
\caption{Hardware and computation flow of axis-shared rasterization.}
\label{fig:ras_array}
\vspace{-5pt}
\end{figure}

\textbf{Inspiration and overview.} We propose axis-shared rasterization to address \textit{Challenge-1} (\secref{sec:challenge_red}). The approach involves three stages: \textit{1)} computing shared terms along the X and Y axes, \textit{2)} broadcasting them to each PE, and \textit{3)} combining them in each PE to obtain the final result. To efficiently implement axis-shared rasterization, our key considerations involve \textit{(i)} simplifying the control logic of the three-stage process, and \textit{(ii)} minimizing or eliminating the additional storage required. For a tile of size $L\times L$, shared-term computation has $O(L)$ complexity, while the combination stage has $O(L^2)$ complexity, with each shared term reused $L$ times. This observation motivates the use of an $L\times L$ array for the combination stage, supported by $L$-sized preprocessing modules for shared-term computation. Following these principles, \figref{fig:ras_array} (top left) shows the design overview. For a $16\times16$ tile, the design includes: \textit{(i)} a $16\times16$ rasterization array, \textit{(ii)} a 16-element X-PE line, and \textit{(iii)} a 16-element Y-PE line. The X-PE line computes X-axis shared terms, while the Y-PE line computes Y-axis shared terms. Each X-PE and each Y-PE broadcast their results to the corresponding rasterization PEs in the same row or column, repeated $16$ times. The PE size is determined by the algorithm \cite{kerbl20233d}, and this broadcast overhead is small enough to implement and meet timing closure requirements.

\textbf{Computation flow and PE structure.} The rasterization array renders each Gaussian and computes its output to a $16\times16$ tile of pixels continuously in each cycle. The Gaussian parameters are provided as input to the X-PE line and Y-PE line, which perform the axes computation, and the outputs from these PE lines are then fed to the rasterization PE array. To eliminate redundant processing of the Gaussian conic matrix parameters (e.g., computing the factor $-\frac{1}{2}$), we directly store the parameter set $\{-\frac{1}{2}a_i, -\frac{1}{2}b_i, c_i\}$, where $i$ denotes the index of each Gaussian. 

The \textit{X-PE line} receives the Gaussian parameters $c_i$ and $-\frac{1}{2}a_i$, which are broadcast to all X-PEs. Internally, each X-PE line generates $16$ x-coordinates ($x_0 \sim x_{15}$), which are automatically incremented by $16$ when shifting to the next tile on the right. With notations shown in \figref{fig:ras_array} (top right), the detailed X-PE structure is illustrated in \figref{fig:ras_array} (right). For simplicity, the x index is omitted (similarly for y in the following description). Each X-PE consists of one adder and three multipliers. One computation branch calculates the term from X-axis ($x$-term) by computing $(x - \mu_i^x)$ and multiplying it by $c_i$, while the other branch squares $(x - \mu_i^x)$ and multiplies it by $-\frac{1}{2}a_i$ to obtain the X-axis quadratic term ($x^2$-term). The \textit{Y-PE line} receives the Gaussian parameter $-\frac{1}{2}b_i$, which is broadcast to each Y-PE. The y-coordinates ($y_0 \sim y_{15}$) are similarly incremented by $16$ when moving vertically between tiles. As shown in \figref{fig:ras_array}  (right), each Y-PE contains one adder and two multipliers. One computation branch produces the Y-axis term ($y$-term) by computing $(y - \mu_i^y)$ directly, while the other squares $(y - \mu_i^y)$ and multiplies it by $-\frac{1}{2}b_i$ to generate the Y-axis quadratic term ($y^2$-term).  

For the \textit{rasterization PE array}, each PE receives vertical inputs from the corresponding X-PE and horizontal inputs from the corresponding Y-PE based on the pixel coordinates. As shown in \figref{fig:ras_array} (bottom), each rasterization PE comprises only one adder and three multipliers. One multiplier is dedicated to multiplying the opacity factor $o_i$ at the final stage, while the remaining MAC unit combines the inputs from the PE lines. Specifically, the multiplier computes the product $x\times y$, and the two adders subsequently add this result to the $x^2$-term and $y^2$-term, respectively.
\textit{Considering synchronization}, the X-PE line initiates one cycle earlier than the Y-PE line, ensuring that the $x$ term and $y$ term arrive simultaneously at each rasterization PE during the first cycle, with the $x^2$-term and $y^2$-term arriving in the second and third cycles, respectively. After executing the exponential operation and multiplying by $o_i$, the computation of $\alpha^i$ is completed, and the result is then fed into the $\alpha$-blending unit. Due to the continuous processing of Gaussians, the computation of $\alpha^{i+1}$ is completed in the subsequent cycle.

\textbf{Overhead analysis.} 
Compared to the GSCore implementation, our axis-shared rasterization requires an additional X-PE line and Y-PE line while significantly simplifying the design of the rasterization PE. To ensure a fair comparison, the MAC cost of the extra PE lines is amortized across the rasterization PE array. \textit{For $\alpha$-computation,} the total number of multipliers is calculated as $16 \times (3 + 2) = 80$ for the X-PE and Y-PE lines and $16 \times 16 \times 2 = 512$ for the rasterization PE array, resulting in a total of $592$. Similarly, the total number of adders is calculated as $2 \times 16 = 32$ for the extra PE lines and $16 \times 16 \times 2 = 512$ for the rasterization PE array, resulting in a total of $544$. When averaged over a $16 \times 16$ rasterization PE array, our axis-shared approach requires only $2.31$ multipliers and $2.13$ adders per PE. Compared with $8$ multipliers and $4$ adders in the GSCore implementation, the counts are reduced by $63\%$ for the $\alpha$-computation stage. \textit{For the complete rasterization process including $\alpha$-blending,} the GSCore design requires $12$ multipliers and $8$ adders per PE. In contrast, our design requires only $6.31$ multipliers and $6.13$ adders per PE after amortization. This corresponds to an overall MAC reduction of approximately $38\%$.

%% file: text/3_algorithm.tex
\section{MLP-based Order-Independent Transmittance}
\label{sec:alg}
\subsection{Algorithmic Motivation}
\label{sec:alg_motivation}
The sorting process is challenging for on-chip deployment, as analyzed in \textit{Challenge-2} (\secref{sec:challenge_sort}). But it is essential for the original 3DGS due to $\alpha$-blending, which depends critically on the relative depth order determined by the camera pose. However, reexamining the blending Equation~\eqref{eq:a_blending}, we observe that the ultimate objective of sorting is to compute the correct transmittance $T_{i}$. The Gaussians preceding the $i_\text{th}$ Gaussian have smaller depth, and each contributes a factor of $1-\alpha_j$ (for $j=0,1,\ldots,i-1$) in the cumulative product. Since each $\alpha$ lies in the interval $(0,1)$, as the depth of the $i_\text{th}$ Gaussian increases, the transmittance $T_{i}$ decreases, thereby serving as a decay factor. This naturally raises the question: \textit{Can we directly compute the decay factor based on depth?}

\begin{figure}[ht]
\vspace{-5pt}
\centering
\includegraphics[width=84mm]{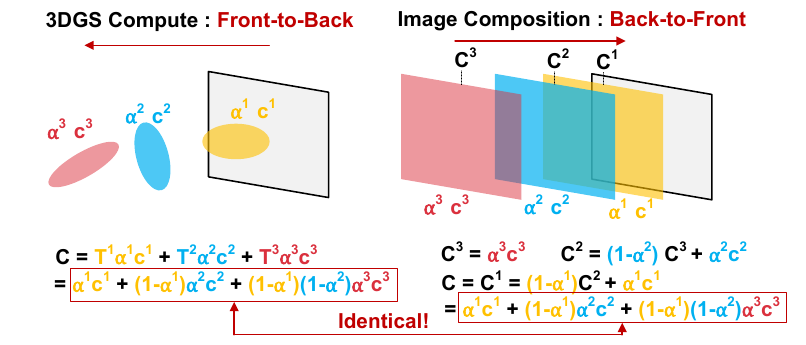}
\vspace{-5pt}
\caption{3DGS $\alpha$-blending is analogous to image composition.}
\label{fig:blending}
\vspace{-5pt}
\end{figure}

\textbf{Inspiration from image composition.} The 3DGS paper~\cite{kerbl20233d} demonstrates that its $\alpha$-blending adopts the NeRF-style volumetric model, while reusing the classic graphics term~\cite{porter1984compositing}. “$\alpha$” stems from the interpolation $\alpha A + (1-\alpha) B$~\cite{smith1995alpha}, corresponding to the ``\textit{over}'' operation of $A$ over $B$. Thus, 3DGS blending is analogous to image composition. \figref{fig:blending} illustrates a three-Gaussian example: 3DGS blends front-to-back, while image compositing is back-to-front. 
In this context, $C_3$, $C_2$, and $C_1$ represent accumulated colors obtained through successive \textit{over} operations with colors and opacities ($c,\alpha$), yielding results identical to 3DGS. This equivalence readily generalizes to any number of Gaussians.

\begin{figure*}[ht]
\centering
\includegraphics[width=176mm]{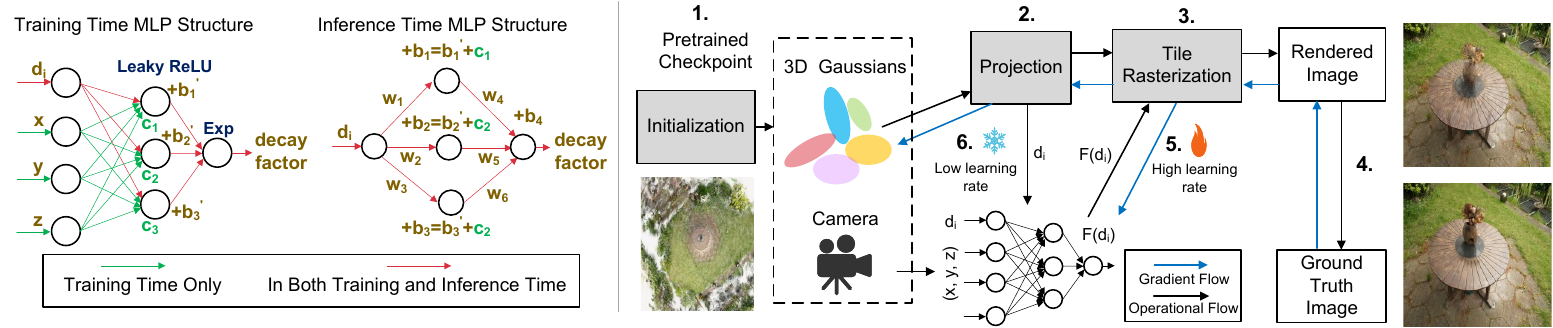}
\vspace{-10pt}
\caption{Neural network structure (left) and the training framework (right).}
\label{fig:training}
\vspace{-10pt}
\end{figure*}

 Because the ``\textit{over}'' operation is non-commutative, depth sorting is computationally expensive. To address this, several order-independent transparency (OIT) techniques have been developed in computer graphics~\cite{carpenter1984buffer, everitt2001interactive, munstermann2018moment, enderton2010stochastic, mcguire2013weighted}. Among them, weighted OIT assigns weights via a monotonic decreasing function of depth $F(d_i)$ and then accumulates colors $C$ accordingly, producing images with negligible quality loss, as formulated in Equation~\eqref{eq:blending}. This success, together with the conceptual similarity between image composition and 3DGS $\alpha$-blending, motivates our exploration of order-independent transmittance (also abbreviated as OIT) for 3DGS.


\begin{equation}
\label{eq:blending}
C = 
\frac{\sum_{i=1}^{n}  F(d_i) \alpha_i c_i}{\sum_{i=1}^{n}  F(d_i) \alpha_i }
\end{equation}

\begin{equation}
\label{eq:view_blending}
C = 
\frac{\sum_{i=1}^{n}  F(d_i, x, y, z) \alpha_i c_i}{\sum_{i=1}^{n}  F(d_i, x, y, z) \alpha_i }
\end{equation}

\textbf{Variables and form of function.} However, weighted OIT relies on manually selected and tuned decay functions $F(d_i)$. Recent work~\cite{hou2024sort} demonstrates that sorting in 3DGS can be approximated by a fixed set of functions. However, their approach is restricted to handcrafted depth-based functions, which limits their ability to model complex dependencies and generalize across diverse scenes. In contrast, we observe that a defining characteristic of 3DGS is its view-dependent synthesis, where camera view critically influences rendering. For different views, even identical Gaussian depths may contribute differently to the final color, and our experiments will further demonstrate the significance of view information in \secref{sec:exp_alg}. To capture this effect, we integrate view information into the input by extending Equation~(\ref{eq:blending}) into Equation~(\ref{eq:view_blending}), where $(x,y,z)$ denotes the normalized view direction vector. This vector, determined only by the camera pose, captures the global viewing ray orientation and is independent of individual Gaussians. Crucially, the interplay between depth and view direction is highly nonlinear and difficult to capture with fixed-form functions. In contrast, a Multi-layer Perceptron (MLP), leveraging its universal regression capability, can model these joint dependencies with high fidelity. Since 3DGS supports rapid per-scene training on desktop or server GPUs, jointly training a small MLP is both practical and computationally efficient, requiring around 30 minutes per scene according to our experiment (\secref{sec:setup}).

\subsection{MLP-based OIT Design and Training Pipeline}
\label{sec:training}

\textbf{Hardware-aware MLP design.} The MLP design decides the  trade-off between algorithmic expressiveness and hardware performance. From a hardware perspective, we aim to reuse the highly parallel MAC units in the rasterization array to accelerate MLP computation, thereby constraining its structure. As shown in \figref{fig:training} (left), the MLP adopts distinct structures for training and inference: green components are active only in training, while red components remain in both phases. During training, the MLP takes as input the Gaussian depth $d_i$ and the view direction $(x,y,z)$, using ground-truth images to learn the decay function. During inference, the view direction is constant across Gaussians for a given camera pose and can therefore be precomputed once. Specifically, its contribution is fused into the MLP’s bias terms: the precomputed values $c_1, c_2, c_3$ are incorporated into the original biases $b_1', b_2', b_3'$, yielding updated biases $b_1, b_2, b_3$. Leaky ReLU ~\cite{xu2020reluplex} with a coefficient of $\tfrac{1}{8}$ serves as the first-layer activation, and an exponential function is used in the output layer. The exponential function is selected for its simplicity and wide dynamic range, while also reusing the exponential unit already present in rasterization. The resulting inference-time MLP is lightweight, containing only $2$ layers and $10$ parameters, and requiring just $6$ MACs, well suited for efficient on-chip deployment.

\textbf{Training pipeline.} Our training framework is illustrated in \figref{fig:training} (right), where black arrows indicate the forward pass and blue arrows denote gradient flow during backpropagation. We initialize the 3D Gaussians with a pre-trained checkpoint from the original sorting-based algorithm (denoted by $1$), which significantly reduces training cost, since co-training Gaussians and the MLP from scratch requires small learning rates and converges slowly. After initialization, the Gaussians are projected into 2D space given the camera pose (denoted by $2$), and their depths are extracted as inputs $d_i$ to the MLP. With sorting eliminated, the MLP outputs $F(d_i)$ enable direct tile rasterization (denoted by $3$), and the rendered image is obtained via Equation~\eqref{eq:view_blending}. The rendered result is then compared against the ground truth to compute the loss (denoted by $4$), following the original setting~\cite{kerbl20233d}. As shown on the rightmost side, early training fails to capture occlusion, producing partially transparent objects. As training progresses, however, the rendered images steadily improve and converge toward the ground truth. 

  \begin{figure*}[ht]
\centering
\includegraphics[width=176mm]{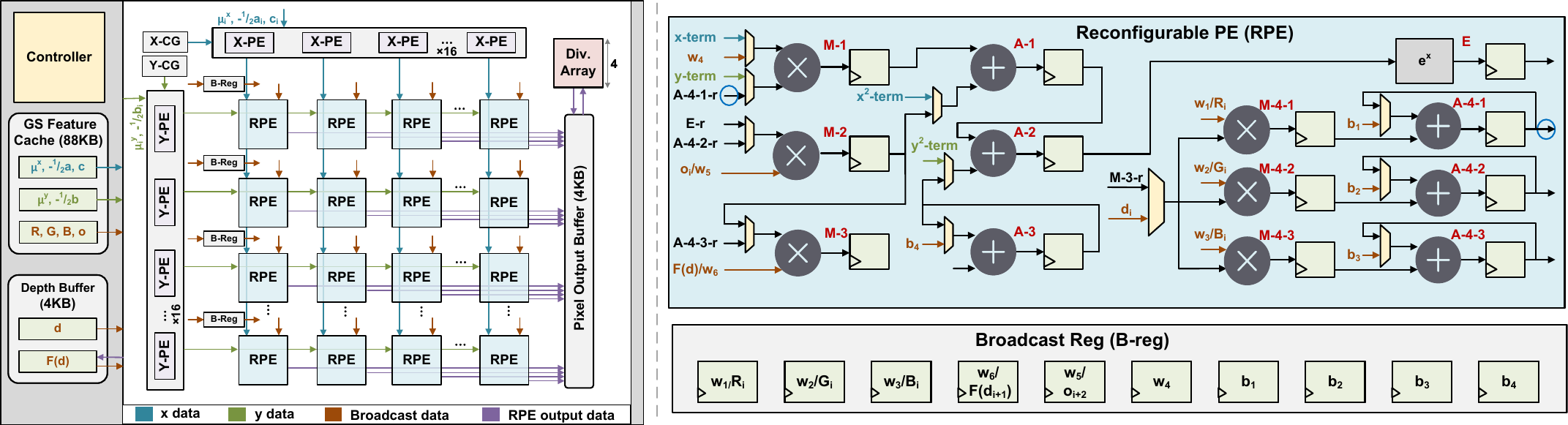}
\vspace{-5pt}
\caption{Unified hardware architecture of rasterization and MLP inference (left), and the structure of modules (right).}
\label{fig:arch}
\vspace{-15pt}
\end{figure*}

The backward process follows the blue paths, where gradients optimize both the 3D Gaussians and the MLP. Their learning rates, however, differ substantially. We apply a scaling factor (e.g., $0.01$) to the original 3DGS learning rate, while assigning a larger learning rate to the MLP. This enables the loss to quickly optimize the MLP (denoted by $5$) while refining the 3D Gaussians more gradually (denoted by $6$). Another key design choice is to disable Gaussian cloning and splitting, as these operations introduce abrupt changes that undermine training stability. Keeping the number of Gaussians constant produces more stable and accurate rendering results. Overall, this framework integrates order-independent transmittance learning into 3DGS, achieving efficient training without explicit depth sorting.

%% file: text/4_hardware.tex
 \section{Architecture Design and Optimization}
 \label{sec:hardware}

 \subsection{Reconfigurable Hardware Design}
 \label{sec:arch}
\textbf{Overview.} \figref{fig:arch} (left) illustrates the unified hardware architecture supporting both rasterization and MLP computation, with arrow notation shown at the bottom left. The architecture consists of two main components: dedicated computation and storage modules. The computation unit features a $16 \times 16$ reconfigurable PE array with per-row broadcast registers, along with X-PE and Y-PE lines (introduced in \secref{sec:axes}). Both PE lines receive $x$ or $y$ coordinates from the coordinate generator (CG), which conducts the tile schedule across the image. For storage, projected Gaussian (GS) features are placed in a feature cache ($9$ parameters per GS), whereas depth values and their decay factors $F(d_i)$ are stored in a separate depth buffer to account for differing on-chip bandwidth demands. Finally, the numerator and denominator from Equation~(\ref{eq:view_blending}) are written to the pixel output buffer and normalized by the division array to produce the final pixel values.

\textbf{Reconfigurable PE.} \figref{fig:arch} (right) shows the reconfigurable PE, consisting of $6$ multipliers, $6$ adders, and one exponential unit, all in FP16 precision. The multipliers and adders (MACs) are organized into two groups of three. In the first group, each MAC operates independently ($M\text{-}\{1\sim3\}$, $A\text{-}\{1\sim3\}$), while in the second group they operate cooperatively ($M\text{-}4\text{-}\{1\sim3\}$, $A\text{-}4\text{-}\{1\sim3\}$). Each MAC is paired with a register to ensure timing closure at high frequency. Some connections are denoted implicitly: the suffix $-r$ marks a register output. For example, $A\text{-}4\text{-}1\text{-}r$ (fourth wire on the left, blue circle) connects to the register output of the $A\text{-}4\text{-}1$ adder (second wire on the right, blue circle) in the same PE. Multiplexer control selects the datapath configuration, switching between rasterization mode (upper) and MLP mode (lower). The workflow and corresponding PE configurations are as follows:

\begin{figure}[ht]
\vspace{-5pt}
\centering
\includegraphics[width=84mm]{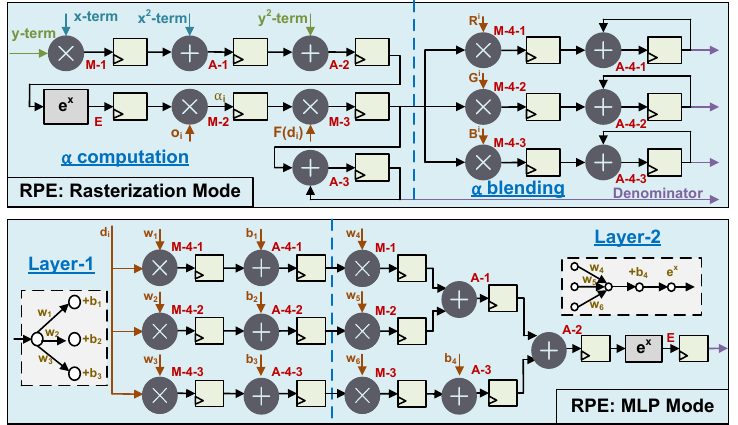}
\vspace{-5pt}
\caption{Reconfigurable PE for rasterization and MLP modes.}\label{fig:rpe}
\end{figure} 


\textbf{Rasterization mode.} In rasterization mode, the PE is configured as shown in \figref{fig:rpe} (top). The PE array processes one GS per cycle for the $16\times16$ tile pixels. The GS-feature buffer supplies GS position ($\mu^{x}_{i}, \mu^y_{i}$) and conic parameters ($-\tfrac{1}{2}a_{i}$, $-\tfrac{1}{2}b_{i}$, $c_i$) to the X-PE and Y-PE lines, where $i$ is the GS index. The unique $o_i$ is loaded into the broadcast register and broadcast $16$ times to all PEs on the same line. Within each PE, $M\text{-}\{1\sim2\}$, $A\text{-}\{1\sim2\}$, and the exponential unit perform the rasterization, as shown in \figref{fig:ras_array}. The X-PE, Y-PE structure and $\alpha$ computation workflow are described in \secref{sec:axes}. The additional step is $\alpha$-blending, defined in Equation \eqref{eq:view_blending}:

GS color features ($R_i, G_i, B_i$) are loaded into the broadcast register, which has $10$ units, $5$ of which are used in rasterization mode (\figref{fig:arch}, right bottom). The decay factor $F(d_i)$, sourced from the depth buffer, is also loaded into the broadcast register. After broadcasting $R_i, G_i, B_i$, and $F(d_i)$ $16$ times to each line, the PEs compute Equation \eqref{eq:view_blending}. Specifically, $M\text{-}3$ multiplies $F(d_i)$ with $\alpha_i$, and $A\text{-}3$ accumulates $F(d_i)\alpha_i$ across Gaussians to form the denominator. Meanwhile, $M\text{-}4\text{-}\{1\sim3\}$ multiplies $F(d_i)\alpha_i$ with $R_i, G_i, B_i$, and $A\text{-}4\text{-}\{1\sim3\}$ accumulates the results across GSs for the RGB channels.


\textbf{MLP mode.} The MLP mode configuration of the PE is illustrated in \figref{fig:rpe} (bottom). In this mode, the PE array processes $16\times16=256$ depth values per cycle, producing $256$ corresponding $F(d)$ outputs. Inputs are read from the depth buffer with high on-chip bandwidth, and the results are written back to the same buffer. The MLP weights reside in the broadcast register, fully utilizing all ten units. As shown, $M\text{-}4\text{-}\{1\sim3\}$ and $A\text{-}4\text{-}\{1\sim3\}$ compute the first layer, including bias addition. The second layer is handled by $M\text{-}\{1\sim3\}$, with $A\text{-}\{1\sim3\}$ performing accumulation and bias addition. The exponential unit is reused to implement the exponential activation function. For simplicity, Leaky ReLU is omitted from the figure. It is implemented via sign detection and a 5-bit integer adder, subtracting $3$ from the FP16 exponent when the input is a normal negative. For subnormal negative or positive values, the input remains unchanged.

 \subsection{Fine-grained Interleaved Pipeline}
 \label{sec:fusion}
   \begin{figure}[ht]
\centering
\includegraphics[width=84mm]{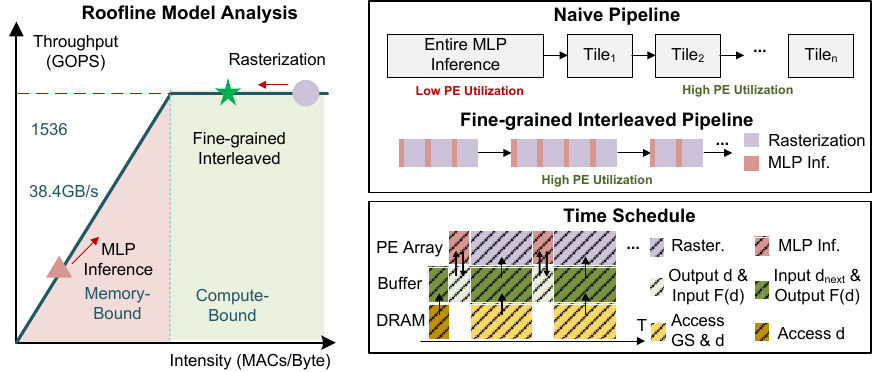}
\vspace{-5pt}
\caption{Roofline model analysis and pipeline comparison.}
\vspace{-5pt}
\label{fig:pipe}

\end{figure} 
\textbf{Memory-bound issue.} Since our MLP inference only relies on the depth of each GS, irrelevant to the tile assignment, it is straightforward to compute the $F(d)$ for the entire GSs, followed by tile-by-tile rasterization, as shown in the naive pipeline of \figref{fig:pipe} (top right).
However, we observe that PE utilization is extremely low during MLP inference. The reason for this is that the operational intensity of MLP inference and rasterization differs significantly. 
For rasterization, each projected Gaussian comprises $9$ parameters and performs $256 \times 6$ MAC operations. In contrast, for MLP inference, $1$ depth parameter only incurs $6$ MAC operations, indicating a nearly $30-$fold difference. 
Given a typical configuration for our architecture, as shown in \figref{fig:pipe} (left), rasterization is compute-bound, whereas MLP inference is heavily memory-bound~\cite{williams2009roofline}. This memory-bound issue also makes deploying our optimized order-independent transmittance algorithm directly to GPUs not an optimal choice. 

\textbf{Optimization.} We propose a fine-grained interleaved pipeline that subdivides each tile into subtiles, as shown in \figref{fig:pipe} (right). The key idea is to overlap rasterization of the current subtile with memory access for MLP inference of the next. Fine-grained subtile processing is required because tiles contain variable numbers of GS while the depth buffer capacity is limited. \figref{fig:pipe} (bottom right) shows how the pipeline maps onto hardware. Starting with the first subtile, DRAM transfers GS depths $d$ to the depth buffer, which then forwards them to the PE array for MLP inference. The resulting $F(d)$ values are written back to the depth buffer. The PE array then rasterizes using $F(d)$ from the depth buffer, while the buffer simultaneously loads depths for the next subtile, overlapping rasterization with memory access. This process repeats, fully hiding depth-access latency except for the first subtile.


 \subsection{Off-chip Access Optimization}
  \label{sec:traj}

\begin{figure*}[ht]
\centering
\includegraphics[width=176mm]{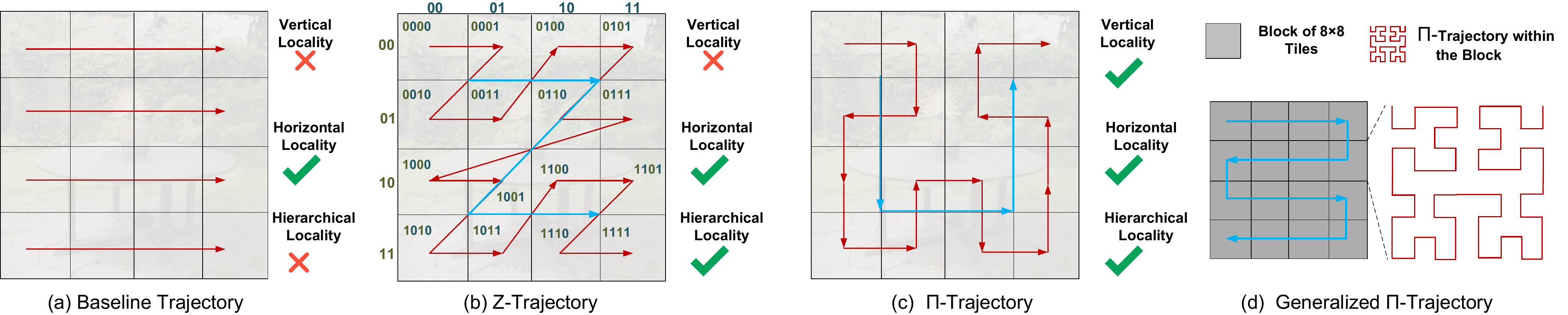}
\vspace{-5pt}
\caption{Comparison of different tile schedule trajectories with $4\times4$ size as the example.}
\label{fig:traj}
\vspace{-10pt}
\end{figure*}

\textbf{Tile schedule.} Our architecture adopts a \textit{GS-feature} cache, where each cache line is tagged using a 28-bit GS ID, and an additional 4 bits record the number of tiles intersected by each GS. The cache prioritizes replacing less important GSs. These 4 bits, together with the 28-bit ID, form a 32-bit aligned storage entry. This design effectively reduces off-chip accesses by leveraging the spatial locality of GSs. However, different tile scheduling trajectories influence cache hit rates. As shown in \figref{fig:traj}(a), the baseline implementation scans tiles in a row-by-row manner, exploiting horizontal locality but lacking vertical and hierarchical locality. For instance, it can reuse GSs intersecting horizontally aligned tiles (e.g., $1\times2$, $1\times3$), but not vertically aligned ones (e.g., $2\times1$, $3\times1$). A slight improvement is achieved using the S-trajectory, which reverses direction at the end of each row. 

Inspired by Morton encoding \cite{morton1966computer}, which interleaves the $y$- and $x$-axis bits (e.g., interleaving $y_1y_0$ with $x_1x_0$ to obtain $y_1x_1y_0x_0$), we schedule tiles in increasing Morton code order, forming a Z-trajectory that better preserves 2D locality, as shown in \figref{fig:traj}(b). However, the diagonal segments of this trajectory may span large spatial distances, causing discontinuities. To mitigate this issue, and inspired by the continuity of Gray code \cite{gray1953pulse}, we modify the Z-trajectory into a ``$\pi$'' trajectory with improved continuity, as shown in \figref{fig:traj}(c). Both Z- and $\pi$-trajectories exhibit hierarchical locality, as indicated by the blue arrow (trajectory of $2\times2$ tiles), making them inherently scalable. This curve corresponds to the Hilbert curve \cite{hilbert1935stetige}, originally proposed for fractal geometry in 1891. We further generalize the design for different 3DGS image resolutions, as shown in \figref{fig:traj}(d). Specifically, the $\pi$-trajectory is applied only within each $8\times8$ tile block, while block-level traversal follows the S-trajectory. For images where the tile count is not divisible by $8$, the remaining tiles are scheduled using a row-wise S-trajectory.

%% file: text/5_evaluation.tex
\section{Evaluation}

\label{sec:eval}
\subsection{Experimental Methodology}
\label{sec:setup}
\textbf{Algorithm Setup.}  
\textit{Datasets and baselines:} Our implementation and GPU-based inference are built on \texttt{Gsplat}, a widely used and efficient 3DGS library \cite{ye2025gsplat}. Following \cite{kerbl20233d}, we evaluate on real-world scenes from the MipNeRF-360 dataset \cite{barron2022mip}, including \textit{garden}, \textit{bicycle}, \textit{stump}, \textit{bonsai}, \textit{counter}, \textit{kitchen}, and \textit{room}. We compare our order-independent transmittance method with the sort-free weight-sum 3DGS algorithm \cite{hou2024sort}. \textit{Evaluation metrics:} Rendering quality is evaluated using three standard metrics: Peak Signal-to-Noise Ratio (PSNR, higher is better), Structural Similarity Index (SSIM, higher is better), and Learned Perceptual Image Patch Similarity (LPIPS, lower is better). \textit{Implementation details:} Training is performed on an NVIDIA RTX 3090 GPU, with checkpoints obtained after $7000$ epochs following \cite{kerbl20233d}. The model is then trained for an additional $10000$ epochs, initialized from these checkpoints. The MLP learning rate is set to $0.005$, while Gaussian learning rates are scaled by $0.01$. Thanks to its lightweight design and the convenient training framework, training is highly efficient, requiring around 30 minutes per scene on an RTX 3090 GPU. Moreover, the model is trained offline only once and can be reused across various edge deployments.



\textbf{Hardware Setup.}  
Our architecture is implemented in SystemVerilog and synthesized with Design Compiler using the TSMC $28$nm CMOS library. Rendering is performed with FP16 arithmetic based on DesignWare IP \cite{synopsys_designware_library}. The design is fully pipelined and operates at $1$~GHz. A DDR5-4800 DRAM with $38.4$~GB/s bandwidth is modeled using Ramulator \cite{kim2015ramulator}, while on-chip SRAM energy and area are estimated with CACTI \cite{muralimanohar2009cacti}. Total energy consumption, including both on-chip and off-chip memory accesses, is obtained using DRAMPower \cite{chandrasekar2012drampower}. Latency and memory traffic are evaluated with a cycle-accurate simulator, cross-validated against RTL simulation results. \tabref{tb:hardware} summarizes the design metrics, and our design occupies $3.85$~mm$^2$ and consumes $1.64$~W. 

\textbf{Baselines.} We benchmark against the NVIDIA Jetson Orin Nano edge GPU~\cite{nvidia_orin_nano} and the desktop-class NVIDIA RTX 3090, demonstrating substantially lower area and power while achieving higher performance, as detailed in \secref{sec:hardware_comp}. We further compare against the state-of-the-art 3DGS accelerators GSCore, MetaSapiens, and GBU~\cite{lee2024gscore,10.1145/3669940.3707227,ye2025gbu}.

\begin{table}[ht]
\vspace{-10pt}
 \centering
\caption{Area and power of our design.}
 \label{tb:hardware}
 \scalebox{0.66}{
\begin{tabular}{cccc}
\hline
Component                & Configuration           & Area \text{[mm$^2$]} & Power \text{[W]} \\ \hline
Reconfigurable PE Array & 16×16 Reconfigurable PE & 2.958   & 1.48    \\
Support Modules & \begin{tabular}[c]{@{}c@{}}X-PE Line + Y-PE line + \\ Coord. Gen. + Div. Array (4 Div.)\end{tabular} & 0.064 & 0.02  \\
On-chip Buffer  & \begin{tabular}[c]{@{}c@{}}GS Feature (88KB) +\\  Output (4KB) + Depth (4KB)\end{tabular}              & 0.826 & 0.14 \\ \hline
Total                   &                         & 3.85    & 1.64    \\ \hline
\end{tabular}
}
\end{table}

\begin{table}[t]
\vspace{-5pt}
\centering
\caption{Overhead comparison between GPUs and our design.}
\label{tb:gpu}
\setlength{\tabcolsep}{3pt}
\renewcommand{\arraystretch}{1.0}
\footnotesize \scalebox{0.8}{
\begin{tabular}{c c c c c c c}
\toprule
Device & Tech. & Area & Power & On-chip SRAM & DRAM Bandwidth & Cores \\
\midrule

\begin{tabular}[c]{@{}c@{}}Jetson Orin Nano\\(8GB)\end{tabular}
& 8 nm 
& 200 
& $\sim$15 W 
& \begin{tabular}[c]{@{}c@{}}$\sim$3 MB\\(L1+L2)\end{tabular}
& 68.2 GB/s 
& \begin{tabular}[c]{@{}c@{}}1024\\CUDA\end{tabular} \\

RTX 3090
& 8 nm 
& 628 
& 350 W 
& \begin{tabular}[c]{@{}c@{}}6 + 10.25 MB\end{tabular}
& 936 GB/s 
& \begin{tabular}[c]{@{}c@{}}10496\\CUDA\end{tabular} \\

Ours 
& 28 nm 
& 3.85 
& 1.64 W 
& 96 KB 
& 38.4 GB/s 
& 256 PEs \\

\bottomrule
\end{tabular}}
\vspace{-10pt}
\end{table}
\subsection{Algorithm Accuracy Evaluation}
\label{sec:exp_alg}
\begin{figure}[ht]
\centering
\includegraphics[width=84mm]{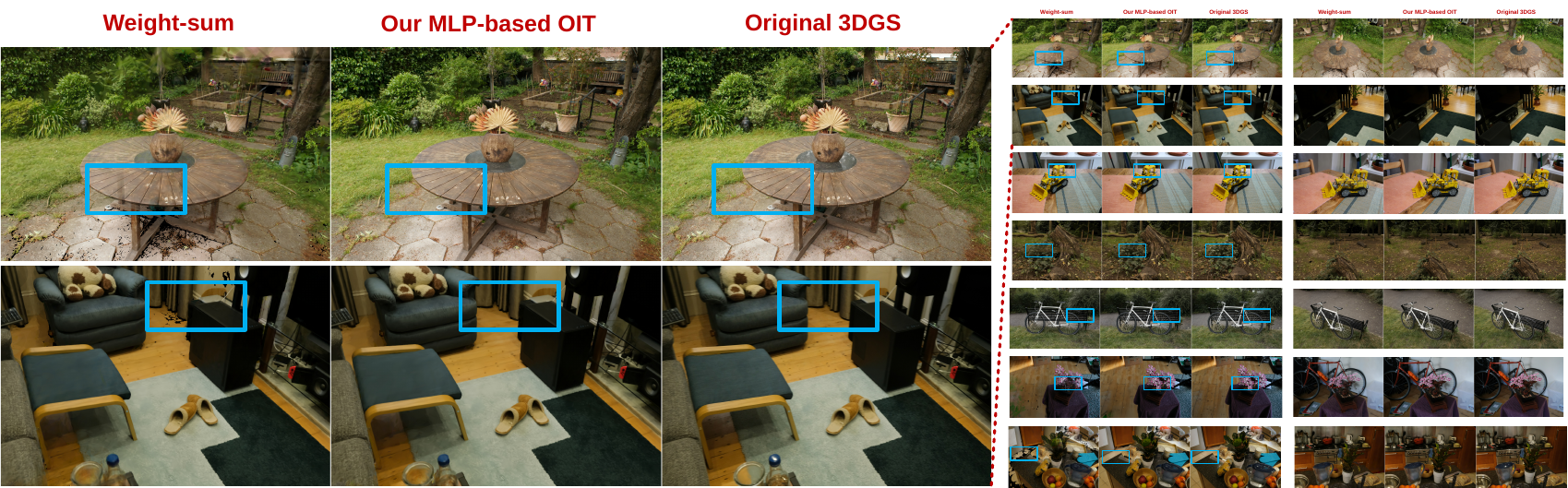}
\vspace{-5pt}
\caption{ Visual comparison among sort-free methods.}
\label{fig:vis}
\vspace{-5pt}
\end{figure}

\textbf{Render quality comparison.} 
Our axis-shared rasterization is numerically identical to the original implementation; therefore, rendering quality differences arise solely from the MLP-based order-independent transmittance (OIT). Table~\ref{tb:img} reports the rendering quality of our MLP-based OIT and compares it with (i) the original 3DGS baseline and (ii) the state-of-the-art sort-free weight-sum rendering method~\cite{hou2024sort}. The weight-sum method in~\cite{hou2024sort} proposes several depth-based weighting functions; we report the best-performing variant, LC-WSR (without view-dependent opacity), for fair comparison. To further analyze the role of view information, we evaluate an ablation version of our model using depth-only input (denoted as OIT+d), while the full model incorporating both depth and view direction is denoted as OIT+d+view. As shown in Table~\ref{tb:img}, incorporating view information yields consistent improvements. PSNR increases from $26.17$ (OIT+d) to $26.90$ (OIT+d+view), accompanied by higher SSIM and lower LPIPS. Overall, our full model achieves a PSNR of $26.90$, with only a minor $0.3$ degradation compared to the original sorted baseline. SSIM remains nearly unchanged, and LPIPS is slightly improved. Compared with prior weight-sum rendering~\cite{hou2024sort}, our approach consistently achieves better results across all three evaluation metrics. \figref{fig:vis} visually compares our method with weight-sum rendering~\cite{hou2024sort} and original 3DGS, as highlighted by the blue boxes, while the weight-sum method exhibits slight blending artifacts in these regions, our method better preserves local occlusion relationships and depth layering, producing results that more closely match the original 3DGS.  

 \begin{table}[hbtp]
 \vspace{-10pt}
 \centering
\caption{image quality comparison with baseline and the sota solution.}
 \label{tb:img}
 \scalebox{0.71}{
\begin{tabular}{l|llllllll}
\hline
Scene        & Bicycle & Bonsai & Counter & Garden & Kitchen & Room   & Stump  & Avg.   \\ \hline
Base. PSNR↑  & 23.71   & 29.66  & 27.14   & 26.30  & 28.86   & 29.21  & 25.62  & 27.21  \\
Weight-sum \cite{hou2024sort}    
             & 23.05   & 26.65  & 24.96   & 24.67  & 26.24   & 28.11  & 24.37  & 25.43  \\
OIT+d        & 23.28   & 27.44  & 26.09   & 26.01  & 26.72   & 28.93  & 24.78  & 26.17  \\
OIT+d+view   & 23.87   & 28.80  & 26.83   & 26.25  & 28.34   & 28.97  & 25.23  & \textbf{26.90}  \\ \hline

Base. SSIM ↑ & 0.6684  & 0.9223 & 0.8782  & 0.8333 & 0.9022  & 0.8930 & 0.7200 & 0.8309 \\
Weight-sum \cite{hou2024sort}    
             & 0.6604  & 0.8772 & 0.8240  & 0.6251 & 0.8692  & 0.8811 & 0.6809 & 0.7700 \\
OIT+d        & 0.6643  & 0.8948 & 0.8668  & 0.8215 & 0.8745  & 0.8916 & 0.6826 & 0.8137 \\
OIT+d+view   & 0.6810  & 0.9081 & 0.8747  & 0.8276 & 0.8986  & 0.8966 & 0.6975 & \textbf{0.8263} \\ \hline

Base. LPIPS↓ & 0.3240  & 0.1623 & 0.2062  & 0.1232 & 0.1272  & 0.2171 & 0.2530 & 0.2017 \\
Weight-sum \cite{hou2024sort}   
             & 0.2667  & 0.2037 & 0.2321  & 0.1930 & 0.1490  & 0.1964 & 0.2504 & 0.2122 \\
OIT+d        & 0.2620  & 0.1785 & 0.1860  & 0.1037 & 0.1451  & 0.1859 & 0.2490 & 0.1869 \\
OIT+d+view   & 0.2454  & 0.1581 & 0.1759  & 0.0969 & 0.1226  & 0.1844 & 0.2339 & \textbf{0.1739} \\ \hline
\end{tabular}}

\end{table}

\textbf{Methodological comparison with related works.}
Weight-sum rendering~\cite{hou2024sort} adopts a monotonic depth-based weighting function. While computationally lightweight, it remains a handcrafted approximation that cannot fully capture complex Gaussian interactions in anisotropic 3DGS scenes. In contrast, our MLP-based OIT provides a data-driven representation. The MLP offers greater expressive capacity than prior methods and flexibly learns an optimized mapping for transmittance estimation. A \textit{key innovation} of our work is the incorporation of view information into order-independent transmittance computation, thereby effectively leveraging the inherently view-dependent nature of 3DGS rendering. Moreover, the interaction between view direction and depth is effectively captured by the MLP, whereas such interactions are difficult to model using handcrafted depth-based functions. Table~\ref{tb:img} further demonstrates that incorporating view information (OIT+d+view) consistently outperforms the variant using only depth information (OIT+d).

\textbf{Advantages over related works.}
\textit{(1) Higher accuracy.}
As shown in Table~\ref{tb:img}, our method consistently exhibits smaller degradation in PSNR and SSIM relative to weight-sum approaches, while maintaining competitive LPIPS scores. 
\textit{(2) Higher training efficiency and practicality.}
Despite being MLP-based, our formulation remains lightweight, and it requires only approximately 30 minutes of additional training per scene within our training framework. In contrast, weight-sum approaches typically require training from scratch with reduced learning rates to ensure stability, resulting in longer convergence times. 
\textit{(3) Lower hardware cost.}
Importantly, our MLP-based OIT is not merely an algorithmic alternative; it is developed from a hardware--algorithm co-design perspective. Although it involves more MAC operations than weight-sum rendering, it reuses the MAC datapath and exponential units already required for rasterization; therefore, MLP inference incurs nearly negligible additional hardware overhead. For example, to match the throughput of our MLP-based OIT within the same accelerator, deploying weight-sum rendering would require additional division units, incurring an extra 0.363\,mm$^2$ area and 341\,mW power, whereas our reconfiguration introduces only 0.147\,mm$^2$ area and 88\,mW power overhead.



\subsection{Hardware Ablation Study}

\begin{figure}[ht]
\vspace{-5pt}
\centering
\includegraphics[width=84mm]{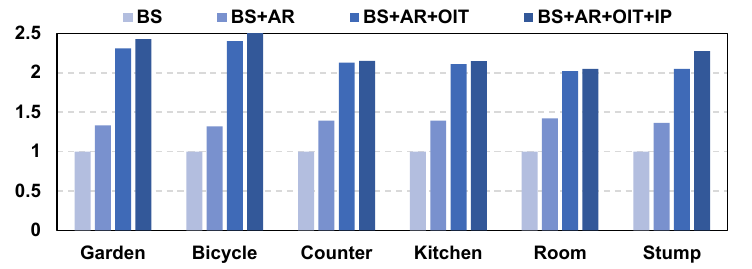}
\vspace{-10pt}
\caption{ Throughput of variants isolating each optimization.}
\label{fig:ab_p}
\vspace{-10pt}
\end{figure}

\textbf{Technique breakdown analysis.}  
We evaluate the impact of each optimization through four design variants:  
(i) \textit{Baseline (BS)}, consisting of a conventional $16\times16$ rasterization array and a $32$-parallel bitonic sorting network \cite{ionescu1997optimizing} following \cite{lee2024gscore}. (ii) \textit{BS+AR}, where the baseline rasterization array is replaced by axis-shared rasterization (AR). (iii) \textit{BS+AR+ OIT}, which applies order-independent transmittance (OIT) to remove the sorting network. It conducts the MLP inference for decay factor computation, replacing the sorting. (iv) \textit{BS+AR+OIT+IP}, which additionally integrates our interleaved pipeline (IP). All variants are compared under the same area budget for fairness. As shown in \figref{fig:ab_p}, axis-shared rasterization improves throughput by $1.37\times$ on average. Adding order-independent transmittance further boosts throughput to $2.16\times$, while the full optimization achieves a $2.27\times$ geometric mean throughput improvement.

\begin{figure}[ht]
\vspace{-10pt}
\centering
\includegraphics[width=84mm]{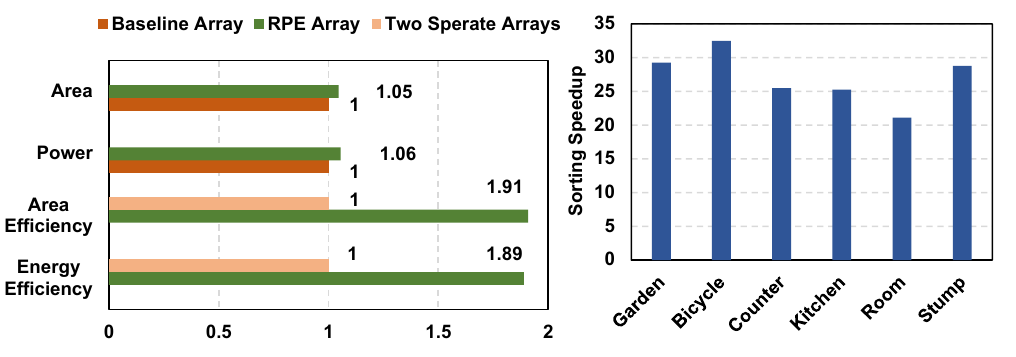}
\vspace{-10pt}
\caption{ Reconfiguration analysis and sorting speedup.}
\label{fig:dce}
\vspace{-5pt}
\end{figure}

\textbf{Reconfiguration analysis and sorting comparison.}  
We compare our reconfigurable design with a baseline PE array dedicated exclusively to rasterization, maintaining the same MAC count but without reconfigurability. Specifically, the baseline array comprises $6$ MUL, $6$ ADD, and $1$ EXP unit per PE, whereas our reconfigurable array contains the same arithmetic units augmented by multiplexers to support reconfiguration. Experimental results in \figref{fig:dce} (left) indicate that reconfiguration incurs only a $5\%$ area overhead and a $6\%$ power overhead. The additional latency overhead is minimal, requiring two extra cycles—one for mode configuration and one for register clearing. Compared with a naive design using separate arrays for MLP inference and rasterization, our architecture delivers $1.91\times$ higher area efficiency (throughput/area) and $1.89\times$ higher energy efficiency (throughput/power), underscoring the effectiveness of the reconfigurable architecture. Relative to the $32$-parallel bitonic sorting network \cite{ionescu1997optimizing} with hierarchical sorting \cite{lee2024gscore}, our reconfigurable PE array for MLP inference achieves a $21.1\times\sim32.4\times$ speedup, as shown in \figref{fig:dce} (right). This substantial speedup, combined with negligible quality degradation, demonstrates the effectiveness of order-independent transmittance as both an efficient and practical solution.

\begin{figure}[ht]
 \vspace{-5pt}
\centering
\includegraphics[width=84mm]{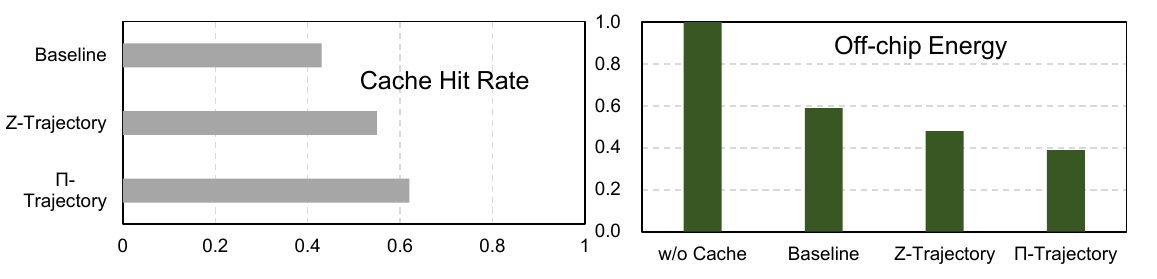}
\vspace{-10pt}
\caption{Cache hit rate and energy comparison.}
\label{fig:cache}
\vspace{-10pt}
\end{figure}

\textbf{Tile schedule trajectory study.} \figref{fig:cache} (left) presents the average cache hit rate across scenes for three tile scheduling methods: the baseline trajectory, the Z-trajectory, and our generalized $\pi$-trajectory tile schedule. The baseline achieves a hit rate of $43\%$, the Z-trajectory $55\%$, and our method improves the hit rate to $62\%$. The corresponding off-chip access energy is shown on the right, with the configuration without cache normalized to $1$. Thanks to the horizontal, vertical, and hierarchical locality utilized, our $\pi$-trajectory tile schedule achieves a $2.56\times$, $1.51\times$, and $1.23\times$ energy saving over no-cache setting, baseline trajectory, and Z-trajectory. 

\subsection{Comparison with Other Implementations}
\label{sec:hardware_comp}

\subsubsection{Comparison with GPUs} 

\begin{figure}[t]
\centering
\includegraphics[width=84mm]{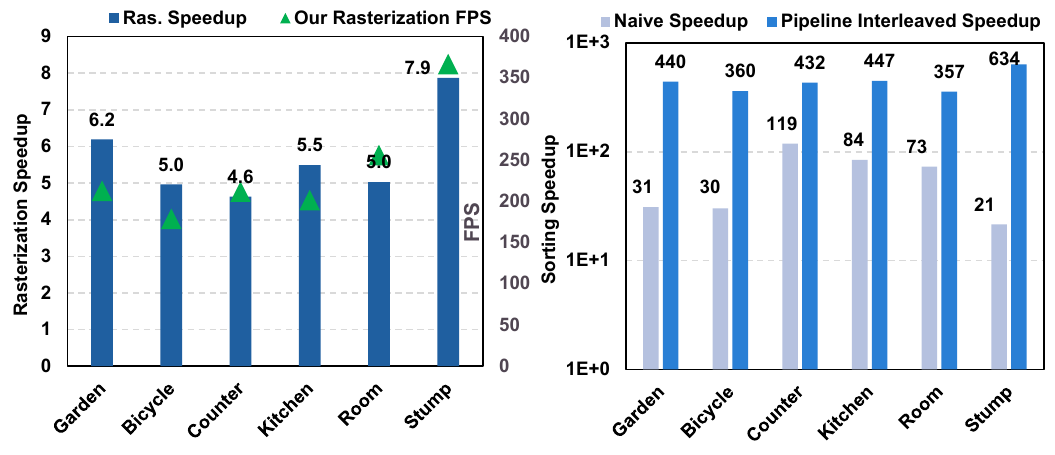}
\vspace{-10pt}
\caption{Speedup of rasterization and sorting over edge GPU.}
\label{fig:ab_speed}
\vspace{-5pt}
\end{figure}

\textit{Rasterization and sorting comparison.} As shown in \figref{fig:ab_speed} (left), our design achieves a rasterization speedup of $4.6\sim7.9\times$ over the edge GPU, with throughput exceeding 150 frames per second (FPS) as indicated on the secondary axis. This speedup stems from axis-shared rasterization, which avoids redundant computations and reduces the MAC count, while our dedicated hardware architecture sustains high parallelism and PE utilization. \figref{fig:ab_speed} (right) compares sorting latency on the edge GPU with that of our MLP-based OIT inference, which replaces the sorting process. It shows that even the naive pipeline achieves a speedup of $21\sim119\times$ over the edge GPU, as order-independent transmittance converts the originally expensive sorting process into a lightweight MAC operation that is efficiently executed by our PE array. By resolving the memory-bound bottleneck, our fine-grained interleaved pipeline further enhances PE utilization, with a speedup over $300\times$.

\begin{figure}[ht]
\vspace{-5pt}
\centering
\includegraphics[width=80mm]{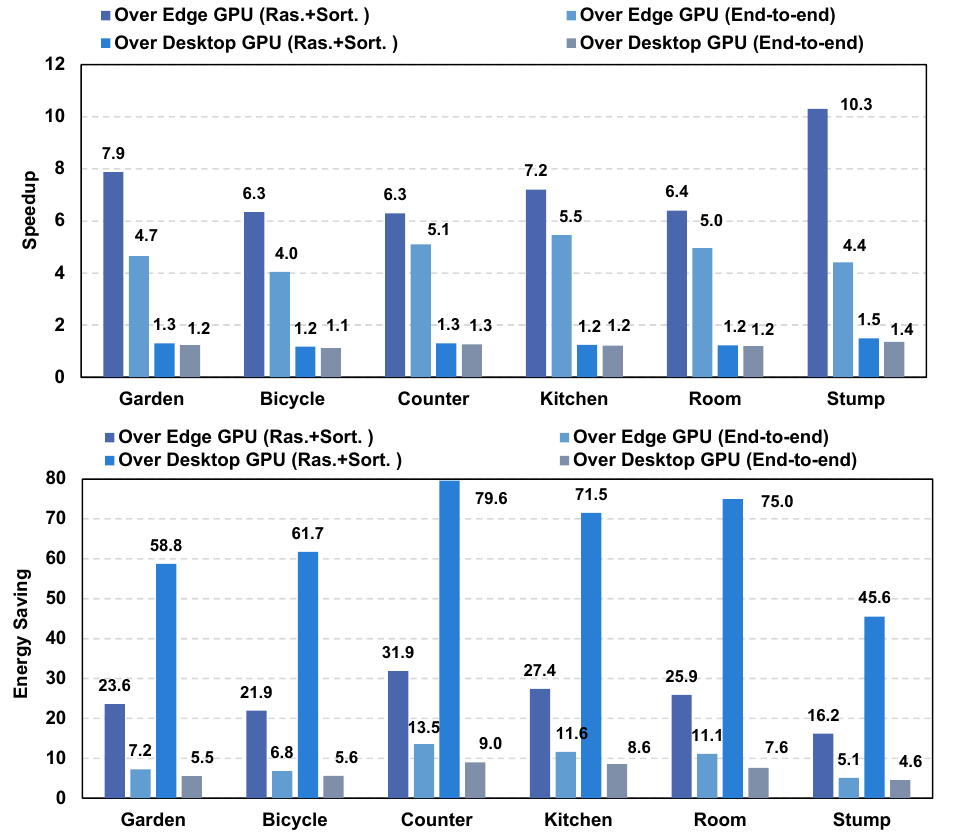}
\vspace{-5pt}
\caption{Overall speedup and energy saving over GPUs.}
\label{fig:gpu_speed}
\end{figure}

\textit{Overall comparison:} We combine our rasterization and sorting optimizations for comprehensive analysis. To simulate practical applications, we report end-to-end performance by executing the Gaussian projection step on the edge GPU, extracting latency and energy consumption, and integrating them into our evaluation. As shown in \tabref{tb:gpu}, our design occupies a much smaller area and is implemented using a less advanced technology node, yet it still achieves substantial speedup and energy savings. \figref{fig:gpu_speed} (top) shows that our combined optimizations achieve a $6.3\sim10.3\times$ speedup over the edge GPU and a $1.2\sim1.5\times$ speedup over the RTX 3090 desktop GPU. For end-to-end speedup, where Gaussian projection dominates latency, our design achieves a $4.0\sim5.5\times$ improvement over the edge GPU and a $1.1\sim1.4\times$ improvement over the desktop GPU. \figref{fig:gpu_speed} (bottom) illustrates that our combined optimizations yield $16.2\times \sim 31.9\times$ energy savings over the edge GPU and $45.6\sim79.6\times$ savings over the desktop GPU. For end-to-end inference, the energy savings are $5.1\sim13.5\times$ over the edge GPU and $4.6\sim9.0\times$ over the desktop GPU. These improvements stem from our dedicated reconfigurable hardware design, MAC reduction via axis-shared rasterization, high parallelism with efficient PE utilization, and the adoption of the order-independent transmittance method.

\subsubsection{Comparison with SOTA accelerators} 

 \begin{figure}[ht]
\centering
\includegraphics[width=84mm]{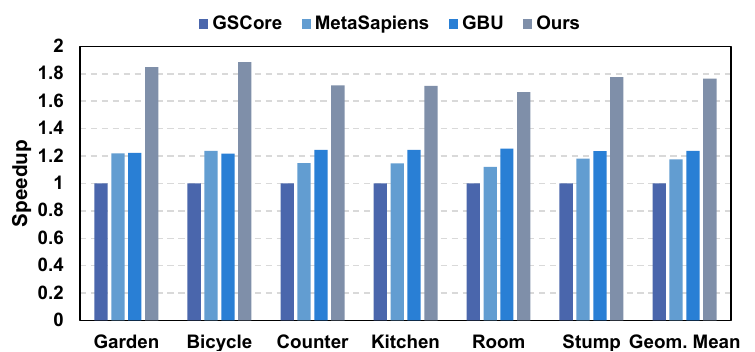}
\vspace{-5pt}
\caption{Speedup over SOTA accelerators.}
\label{fig:comp_gs}
\vspace{-10pt}
\end{figure} 

We compare our design with GSCore \cite{lee2024gscore}, GBU \cite{ye2025gbu}, and MetaSapiens \cite{10.1145/3669940.3707227}, with a focus on the sorting and rasterization stages. All designs are based on a $28$ nm process and operate at $1$ GHz. To ensure fairness, all designs are normalized to the same area budget\footnote{Since GSCore and MetaSapiens propose techniques to reduce the Gaussian count, and these techniques are compatible with our design. To guarantee fairness, we assume that all designs employ the same Gaussian reduction technique as GSCore.}. As shown in \figref{fig:comp_gs}, taking GSCore as the baseline, our design achieves a $1.67\sim1.88\times$ speedup, enabled by the dedicated axis-shared rasterization and order-independent transmittance. MetaSapiens \cite{10.1145/3669940.3707227} mitigates pipeline imbalance through tile merging and an incremental pipeline, but at the cost of additional buffers and complex control, while still leaving rasterization redundancy unaddressed. Our design achieves a $1.49\sim1.52\times$ speedup over MetaSapiens. Compared with GBU, our design achieves a $1.33\sim1.55\times$ speedup. Although GBU also reduces rasterization MACs through sequential computation, it lacks sorting optimization. Furthermore, sequential computation introduces pixel dependencies that limit parallelism scalability, leading to additional overhead.

\subsection{GPU-implemented Optimization Evaluation}
\subsubsection{GPU-implemented axis-shared rasterization} \textbf{Implementation.} To further clarify the benefits of our co-designed accelerator, we implement axis-shared rasterization on an NVIDIA RTX 3090 GPU. Axis-shared rasterization consists of three stages: shared-term computation, broadcast, and combination, as described in \secref{sec:axes}. To map this structure onto the GPU, we assign one $16\times16$ thread block to each tile. Within each block, we introduce an explicit shared-term computation stage, where threads collaboratively compute the X-axis and Y-axis shared terms. These intermediate results are stored in shared memory. After synchronization, all threads reuse the shared terms to perform the combination stage in parallel.

  \begin{figure}[ht]
\vspace{-10pt}
\centering
\includegraphics[width=80mm]{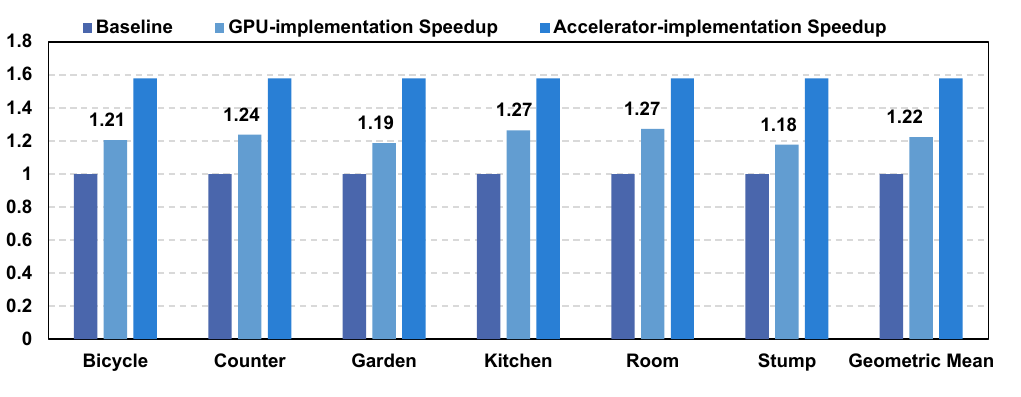}
\vspace{-10pt}
\caption{The effect of GPU-implemented axis-shared optimization.}
\label{fig:gpu_speedup}
\vspace{-5pt}
\end{figure}

\figref{fig:gpu_speedup} shows the performance improvement of a GPU-based implementation of axis-shared rasterization across scenes. The baseline is the original \texttt{Gsplat} implementation~\cite{ye2025gsplat}. For reference, we include the performance of our accelerator implementation (right), which serves as an architectural upper bound. Because axis-shared rasterization reduces MAC count and hardware area, we normalize speedup by comparing latency under an equivalent area budget. The GPU implementation achieves a geometric mean speedup of $22\%$, substantially lower than the nearly $60\%$ speedup delivered by our dedicated accelerator. Because rasterization is a non-GEMM workload, it cannot effectively utilize Tensor Cores and is therefore executed on CUDA cores. Compared with our accelerator, the GPU mapping exhibits limitations in both computational structure and memory behavior.

\textbf{Computational limitations.}
\textit{(i)} Axis-shared rasterization consists of an $O(L)$ shared-term stage followed by an $O(L^2)$ combination stage. This imbalance leads to substantial thread underutilization during the $O(L)$ stage. In contrast, our accelerator exploits the $O(L)$ edge and $O(L^2)$ area relationship of the PE array, mapping shared-term computation to PE lines and the combination stage to the full array. This spatial decomposition enables near-full utilization across both stages.
\textit{(ii)} The GPU MAC configuration cannot fully accommodate the requirements of rasterization. The shared-term stage requires more multiplications than additions, whereas the combination stage involves addition-before-multiplication patterns that deviate from the standard fused multiply–add (FMA) pipeline. GPUs employ balanced multiplier–adder ratios and fixed FMA pipelines, making such irregular arithmetic sequences inefficient. Moreover, frequent exponential operations further increase computational pressure. In contrast, our accelerator adopts a dedicated datapath, allowing the reduced MAC count to translate directly into area and latency savings.

\textbf{Memory limitations.}
\textit{(i)} The shared terms must be written to and read from shared memory between stages, incurring non-negligible synchronization and access overhead. In contrast, our accelerator fully fuses the three stages into a register-to-register datapath without intermediate buffer or memory accesses. \textit{(ii)} The GPU-based axis-shared rasterization trades additional storage for reduced MAC count. Because existing 3DGS GPU kernels already heavily utilize register files and shared memory, the additional reuse of shared terms increases register and shared-memory pressure, raising the risk of register spilling. In contrast, our dedicated accelerator pipeline carefully allocates registers, enabling shared terms to be produced and consumed in place, without extra storage. Overall, these computational and memory constraints highlight the limitations of deploying axis-shared rasterization on GPUs, further motivating a dedicated accelerator design.

\subsubsection{GPU-implemented MLP-based OIT} 
\begin{figure}[!t]
\centering
\includegraphics[width=80mm]{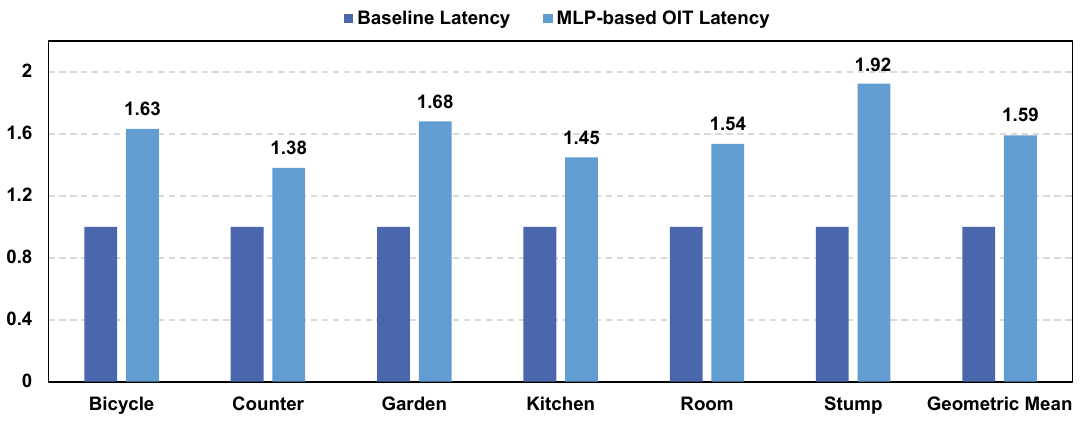}
\vspace{-10pt}
\caption{Latency comparison between baseline sorting and MLP-based OIT.}
\label{fig:gpu_oit}
\vspace{-10pt}
\end{figure}

The MLP-based OIT is also implemented on an NVIDIA RTX 3090 GPU using cuBLAS~\cite{nvidia_cublas}, and its latency is compared against the Radix sorting~\cite{merrill2010revisiting} implementation in \texttt{Gsplat}~\cite{ye2025gsplat} as the baseline. As shown in \figref{fig:gpu_oit}, the GPU-implemented MLP-based OIT is slower, and exhibits a geometric mean latency of $1.59\times$ that of the baseline sorting approach. Although our MLP is extremely lightweight, its inference on the GPU becomes memory bound. As discussed in \secref{sec:fusion}, the small MLP exhibits low arithmetic intensity, thereby limiting throughput under the GPU execution model. As a result, the theoretical arithmetic simplicity of the MLP does not translate into practical latency reduction on the GPU.

In our accelerator, the MLP-based OIT solves the challenge of sorting on edge devices, as described in \secref{sec:challenge_sort}. It is mapped onto the same unified PE array through a reconfigurable design, incurring negligible additional hardware overhead while eliminating inter-stage pipeline imbalance. Furthermore, the fine-grained interleaved pipeline effectively hides memory latency and mitigates the memory-bound bottleneck. These architectural optimizations collectively enable substantial speedup over the baseline GPU sorting approach, as reported in \secref{sec:hardware_comp}, further demonstrating the motivation of our dedicated accelerator design.

\subsection{Applicability to Dynamic Scenes}

{Although our design primarily targets static 3DGS rendering, it is also meaningful to study the applicability of order-independent transmittance (OIT) to dynamic scenarios. We extend OIT from 1D image composition to 3D Gaussian splatting by incorporating view information, a natural question is whether it remains effective for dynamic scenes. To investigate this, we evaluate our MLP-based OIT on the Neu3D dataset~\cite{li2022neural}, a widely used benchmark for real-world dynamic view synthesis, featuring high-resolution sequences (2704×2028) and spanning 10 seconds (300 frames). We adopt 4DGS~\cite{yang2023real} as the baseline for dynamic scene modeling. 4DGS models dynamic scenes using 4D primitives that generate standard 3D Gaussians at each timestamp, making the per-frame rendering pipeline identical to that of static 3DGS. To account for temporal variations, we update the weights of the 10-parameter MLP every 30 frames. For a typical 300-frame sequence, this produces 10 independent sets of MLP parameters covering the entire sequence.}


 \begin{table}[hbtp]
 \centering

\caption{evaluation of our oit applied to dynamic scenes.}
 \vspace{-5pt}
 \label{tb:dynamic}
 \scalebox{0.8}{
\begin{tabular}{c|cc|cc|cc|ll}
\hline
 & \multicolumn{2}{c|}{Cook Spinach} & \multicolumn{2}{c|}{Cut Beef} & \multicolumn{2}{c|}{Flame Steak} & \multicolumn{2}{c}{Average} \\
         & PSNR  & SSIM   & PSNR  & SSIM   & PSNR  & SSIM   & \multicolumn{1}{c}{PSNR} & \multicolumn{1}{c}{SSIM} \\ \hline
Baseline & 32.88 & 0.9572 & 32.75 & 0.9575 & 32.78 & 0.9552 & \textbf{32.80}                    & \textbf{0.9566}                   \\
Our OIT  & 32.46 & 0.9566 & 32.29 & 0.9571 & 32.31 & 0.9550 & \textbf{32.35}                    & \textbf{0.9562}                   \\ \hline
\end{tabular}}
 \vspace{-5pt}

\end{table}

As shown in Table~\ref{tb:dynamic}, our MLP-based OIT maintains high fidelity, with only a $0.45$ PSNR drop compared with the baseline. This result indicates that each localized MLP can effectively capture evolving occlusion relationships within its time window. The additional overhead is minimal, requiring only $10 \times 10$ extra parameters, making the approach well suited for edge deployment. Looking forward, dynamic scene rendering is expected to involve longer sequences and richer temporal variations. While our results demonstrate the feasibility of the proposed method for dynamic scenes, extending it to longer sequences or scenes with more drastic geometric changes may require further adaptation, which we leave for future work.


%% file: text/7_related_work.tex
\section{Related Work}

\textbf{Efficient 3D Gaussian Splatting algorithm.} Recent studies aim to accelerate rendering or reduce memory overhead in 3DGS primarily by pruning Gaussian counts \cite{fan2024lightgaussian,niemeyer2024radsplat,fang2024mini,girish2024eagles,ye20243d,fang2024mini2}, while acceleration of the sorting process remains relatively underexplored. A recent work~\cite{hou2024sort} proposes monotonically decreasing depth-based functions to enable nearly sort-free rendering, demonstrating that learnable functions can replace traditional sorting. However, this method relies solely on depth input and manually designed function forms, limiting expressiveness and generalization across diverse scenes. A detailed discussion was provided in \secref{sec:exp_alg}.


\textbf{3D Gaussian Splatting accelerators.} To meet the real-time rendering demands of edge devices, several specialized hardware accelerators have recently been proposed \cite{ye2025gbu,lee2024gscore,10.1145/3669940.3707227,pei2025gcc3dgsinferencearchitecture,11043574, 10993091}. GBU \cite{ye2025gbu} introduces a rasterization module for edge GPUs designed to reduce MAC operations. However, our approach fundamentally differs in both motivation and implementation. GBU reduces MACs through spatial transformation and eigenvalue decomposition, whereas our method identifies computational redundancy and introduces axis-shared computation to eliminate it, resulting in a distinct hardware architecture. In terms of parallelism, GBU performs differential computations row-wise by reusing consecutive pixels, which introduces dependencies that limit intra-tile parallelism. In contrast, our method enables fully parallel pixel processing within each tile, substantially improving scalability and throughput. GSCore \cite{lee2024gscore} streamlines the Gaussian splatting pipeline by reducing the number of Gaussians assigned to tiles using a shape-aware intersection test and introduces hierarchical sorting through a sorting network, thereby accelerating both rasterization and sorting. MetaSapiens \cite{10.1145/3669940.3707227} leverages efficiency-aware pruning and the low visual acuity of the human periphery to relax rendering quality, improve rendering speed, and accelerate foveated rendering. These Gaussian count reduction techniques \cite{lee2024gscore, 10.1145/3669940.3707227} are compatible with our design, but our approach diverges in two key aspects. First, we identify inherent redundancy in rasterization and propose axis-shared rasterization, supported by a highly parallel architecture that reduces the MAC count without sacrificing generality. Second, we optimize the sorting process through algorithm–hardware co-design: sorting is replaced by order-independent transmittance at the algorithmic aspect, while our reconfigurable array with an interleaved pipeline reduces MLP inference latency to a negligible level. In addition, several accelerators focus on domain-specific 3DGS applications. For example, GauSPU~\cite{wu2024gauspu} introduces a co-processor for 3DGS-based SLAM, while GsArch~\cite{he2025gsarch} alleviates memory bottlenecks to improve training efficiency. Lumina~\cite{10.1145/3695053.3731003} targets moving-view scenarios and reduces sorting and rasterization costs by reusing results across consecutive frames.

\textbf{MAC reduction in traditional splatting.}
FastSplat~\cite{huang2000fastsplats} reduces MAC operations via differential computation, reusing intermediate results between consecutive pixels and updating only incremental terms. While effective for sequential software implementations, this formulation introduces inter-pixel dependencies that constrain spatial parallelism and complicate scalable hardware mapping. In contrast, our axis-shared rasterization removes redundancy by precomputing and sharing common terms along the X- and Y-axes within each tile, without introducing sequential dependencies. This design enables full spatial parallelism across processing elements while reducing redundant arithmetic operations. Moreover, FastSplat targets traditional splatting, whereas our work focuses on anisotropic Gaussian rasterization in modern 3DGS pipelines with different computational structures.

\textbf{GPU-based 3DGS acceleration.} Orthogonal to dedicated 3DGS accelerator design, many studies focus on GPU-based acceleration by employing efficient scheduling strategies and CUDA kernel optimization~\cite{arc,tao2025gscachegscacheinferenceframework,yuan2025efficientdifferentiablehardwarerasterization,feng2024flashgsefficient3dgaussian,hanson2025speedysplatfast3dgaussian,gui2025balanced3dgsgaussianwiseparallelism,liao2025tcgsfastergaussiansplatting}. For instance, FlashGS \cite{feng2024flashgsefficient3dgaussian} and Speedy-Splat \cite{hanson2025speedysplatfast3dgaussian} develop more accurate Gaussian-intersection detection primitives, while Balanced 3DGS~\cite{gui2025balanced3dgsgaussianwiseparallelism} reduces thread idling and mitigates load-imbalance issues through dynamic workload distribution. Additionally, TC-GS \cite{liao2025tcgsfastergaussiansplatting} transforms the $\alpha$-computation into matrix multiplication and utilizes otherwise idle Tensor Cores to accelerate the rendering pipeline. GauRast \cite{li2025gaurast} augments existing GPU triangle rasterizers to support Gaussian-based rendering. By leveraging the structural similarities between traditional graphics pipelines and 3DGS, it achieves substantial energy savings and throughput gains for edge deployment. 


%% file: text/6_conclusion.tex
\section{Conclusion}

This work presents an architecture–algorithm co-design to enable real-time 3D Gaussian Splatting on resource-cons-
trained platforms. We identify two overlooked challenges, computational redundancy in rasterization and scalability and imbalance issues in sorting, and address them through novel solutions. First, the proposed \textit{axis-shared rasterization} eliminates redundant computations by sharing intermediate terms across processing elements, reducing MAC operations by $38\%$. Second, our \textit{order-independent transmittance} method bypasses explicit sorting with a lightweight MLP, mitigating pipeline bottlenecks while maintaining rendering quality. Finally, we develop a unified and reconfigurable hardware accelerator that sustains high utilization across rasterization and MLP inference. This work provides a practical and architecture-friendly foundation for efficient 3DGS acceleration, enabling broader adoption of real-time and efficient rendering on resource-constrained systems.